\documentclass[aps,twocolumn]{revtex4}
\usepackage{epsf}% Include figure files
\begin{document}

%\preprint{cond-mat/0305xxx}

\title{Weak magnetism and non-Fermi liquids near heavy-fermion critical points}

\author{T. Senthil}
\affiliation{Department of Physics, Massachusetts Institute of
Technology, Cambridge MA 02139}

\author{Matthias Vojta}
\affiliation{Institut f\"ur Theorie der Kondensierten Materie,
Universit\"at Karlsruhe, Postfach 6980, 76128 Karlsruhe, Germany}

\author{Subir Sachdev}
\affiliation{Department of Physics, Yale University, P.O. Box
208120, New Haven CT 06520-8120}

\date{\today}

\begin{abstract}

This paper is concerned with the weak-moment magnetism in
heavy-fermion materials and its relation to the non-Fermi liquid
physics observed near the transition to the Fermi liquid. We
explore the hypothesis that the primary fluctuations responsible
for the non-Fermi liquid physics are those associated with the
destruction of the large Fermi surface of the Fermi liquid.
Magnetism is suggested to be a low-energy instability of the
resulting small Fermi surface state. A concrete realization of
this picture is provided by a fractionalized Fermi liquid state
which has a small Fermi surface of conduction electrons, but also
has other exotic excitations with interactions described by a
gauge theory in its deconfined phase. Of particular interest is a
three-dimensional fractionalized Fermi liquid with a spinon Fermi
surface and a U(1) gauge structure. A direct second-order
transition from this state to the conventional Fermi liquid is
possible and involves a jump in the electron Fermi surface volume.
The critical point displays non-Fermi liquid behavior.
A magnetic phase may develop from a spin density wave instability of the
spinon Fermi surface. This exotic magnetic metal may have a weak
ordered moment although the local moments do not participate in
the Fermi surface. Experimental signatures of this phase and
implications for heavy-fermion systems are discussed.

\end{abstract}

\maketitle

%\begin{multicols}{2}

\section{Introduction}
\label{intro}

The competition between the Kondo effect and inter-moment exchange
determines the physics of a large class of materials which have
localized magnetic moments coupled to a separate set of conduction
electron~\cite{doniach}. When the Kondo effect dominates, the
low-energy physics is well described by Fermi liquid theory (albeit
with heavily renormalized quasiparticle masses).  In contrast when
the inter-moment exchange dominates, ordered magnetism typically
results.

A remarkable experimental property of such magnetic states is that
the magnetism is often very weak -- the ordered moment per site is
much smaller than the microscopic local moment that actually
occupies each site. The traditional explanation of this feature is
that the magnetism arises out of imperfectly Kondo-screened local
moments. In other words, the magnetism is to be viewed as a spin
density wave that develops out of the parent heavy Fermi liquid
state. We will henceforth denote such a state as SDW. Clearly a
SDW state may be a small moment magnet.

A different kind of magnetic metallic state is also possible in
heavy-fermion materials where the moments order at relatively
large energy scales, and simply do not participate in the Fermi
surface of the metal. In such a situation, the saturation moment
in the ordered state would naively be large, {\em i.e.\/}, of
order the atomic moment.

Often the distinction between these two kinds of magnetic states
can be made sharply: the two Fermi surfaces in the two states may
have different topologies (albeit, the same volume modulo the
volume of the Brillouin zone of the ordered state), so that they
cannot be smoothly connected to one another.

In recent years, a number of experiments have unearthed some
fascinating phenomena near the zero temperature ($T$) quantum
transition between the heavy-fermion liquid and the magnetic
metal. In particular, many experiments do not fit easily
\cite{piers1,stewart,HvL,schroeder} into a description in terms of
an effective Gaussian theory for the spin density wave
fluctuations, renormalized self-consistently by quartic
interactions \cite{maki,hertz,moriya,lonzarich,millis}. This
theory makes certain predictions on deviations from Fermi liquid
behavior as the heavy Fermi liquid state approaches magnetic
ordering induced by the condensation of the spin density wave
mode; those predictions are, however, in disagreement with
experimental findings. This conflict raises the possibility that
the magnetic state being accessed is not in the first category
discussed above: a SDW emerging from a heavy Fermi liquid. Rather,
it may be the second kind of magnetic metal where the local
moments do not participate at all in the Fermi surface. In other
words, the experiments suggest that the Kondo effect (crucial in
forming the Fermi liquid state)  is itself suppressed on
approaching the magnetic state.

This proposal clearly raises several serious puzzles.  How do we
correctly describe the non-Fermi liquid physics near the
transition? If this non-Fermi liquid behavior is accompanied by
the suppression of the Kondo effect, how do we reconcile it with
the weak moments found in the magnetic state? The traditional
explanation for the weak magnetism is apparently in conflict with
the picture that the Kondo effect and the resultant heavy Fermi
liquid state are destroyed on approaching the magnetic state. In
other words, the naive expectation of a large saturation moment in
a magnetic metal where the local moments do not participate in the
Fermi surface must be revisited.

The weakness of the ordered moment in the magnetic state may be
reconciled with the apparent suppression of the Kondo effect if we
assume that there are strong quantum fluctuations of the spins
that reduce their moment. Such strong quantum effects may appear
to be unusual in three-dimensional systems, but may be facilitated
by the coupling to the conduction electrons (even if there is no
actual Kondo screening). In this paper we study specific states
where such quantum fluctuations have significantly reduced the
ordered moment (or even caused it to vanish), and the evolution of
such states to the heavy Fermi liquid.

We begin with several general pertinent observations. First,
consider the heavy Fermi liquid state. This Fermi liquid behavior
is accompanied by a Fermi surface which, remarkably, satisfies
Luttinger's theorem only if the local moments are included as part
of the electron count. (Such a Fermi surface is often referred to
as the ``large Fermi surface'', and we will henceforth refer to
such a phase as FL). The absorption of the local moments into the
Fermi volume is the lattice manifestation of the Kondo screening
of the moments. We take as our starting point the assumption that
the Kondo effect becomes suppressed on approaching the magnetic
state. What then happens to the large Fermi surface?

In thinking about the resulting state theoretically, it is
important to realize that once magnetic order sets in, there is no
sharp distinction between a large Fermi volume which includes the
local moments, and a Fermi volume that excludes the local moments
-- the latter is often loosely referred to as ``small''. This is
because the Fermi volumes can only be defined modulo the volume of
the Brillouin zone, and the onset of magnetic order at least
doubles the unit cell and hence at least halves the Brillouin zone
volume. (There can, however, be a distinction between the Fermi
surfaces topologies in the two situations.)

In this paper we will take the point of view that the primary
transition involves the destruction of the large Fermi surface,
and that the resulting small Fermi surface state has a distinct
physical meaning even in the absence of magnetic order. The
magnetic order will be viewed as a low-energy instability of the
resulting state in which the local moments are not to be included
in the Fermi volume.

Evidence in support of this point of view exists. In the
experiments the non-Fermi liquid behavior extends to temperatures
well above the Neel ordering temperature even far away from the
critical point. This suggests that the fluctuations responsible
for the non-Fermi liquid behavior have very little to do with the
fluctuations of the magnetic order parameter. Some further support
is provided by the results of inelastic neutron scattering
experiments that apparently see critical behavior at a range of
wave-vectors including (but not restricted to) the one associated
with magnetic ordering in the magnetic metal \cite{schroeder}.
Finally, there even exist materials in which the non-Fermi liquid
features persist into the magnetically ordered side -- this is
difficult to understand if the non-Fermi liquid physics is
attributed to critical fluctuations of the magnetic order
parameter.

Conceptually, as we asserted above, it pays to allow for the
possibility of a non-magnetic state in which the suppression of
the Kondo effect removes the local moments from the Fermi volume,
resulting in a ``small Fermi surface'', even though such a state
may not actually be a ground state in the system of interest. In
our previous work \cite{ssv1} we argued that such states do exist
as ground states of Kondo lattice models in regular
$d$-dimensional lattices, and that the violation of Luttinger's
theorem in such a state was intimately linked to the presence of
neutral $S=1/2$ and $S=0$ excitations induced by topological order
(see also Appendix~\ref{oshi}): we dubbed such ground states
FL$^*$.

Clearly, it is worthwhile to explore metallic magnetic states that
develop out of such FL$^*$ states (just as the usual SDW state
develops out of the Fermi liquid). Such states, which we will
denote SDW$^*$, represent a third class of metallic magnetic
states distinct from both the conventional SDW and the
conventional local-moment metal described above. As we will see,
in such magnetic states the local moments do not participate in
the Fermi surface. Nevertheless they may have a weak ordered
moment. Thus these states offer an opportunity for resolution of
the puzzles mentioned above. The properties and the evolution of
such states, and their parent FL$^*$ states, to the Fermi liquid
will be the subject of this paper. The SDW$^*$ states inherit
neutral spin $S=1/2$ spinon excitations and $S=0$ ``gauge''
excitations from the FL$^*$ states, which will be described more
precisely below; these excitations coexist with the magnetism and
the metallic behavior. The experimental distinction between the
SDW and SDW$^*$ states is however subtle, and will also be
described in this paper. (The FL and FL$^*$ states can be easily
distinguished by the volumes of the Fermi surfaces, but this
distinction does not extend to the SDW and SDW$^*$ states.)

We emphasize that a wide variety of heavy-fermion materials
display non-Fermi liquid physics in the vicinity of the onset of
magnetism that is, to a considerable extent, universal. However,
the detailed behavior at very low temperature appears to vary
across different systems. In particular, in some materials a
direct transition to the magnetic state at very low temperature
does not occur (due for instance to intervention of a
superconducting state). In other materials, such a direct
transition does seem to occur at currently accessible
temperatures. In view of this, we will not attempt to predict the
detailed phase diagram at ultra-low temperatures. We focus instead
on understanding the universal non-Fermi liquid physics not too
close to the transition and its relation to the magnetic state.

\subsection{Summary of results}
\label{summary}

Our analysis is based upon non-magnetic translation-invariant
states that have a small Fermi surface (FL$^*$), and the related
transitions to the heavy Fermi liquid (FL). As we showed
previously \cite{ssv1}, the FL$^*$ state has a Fermi surface of
long-lived electron-like quasiparticles whose volume does not
count the local moments. The local moments are instead in a state
adiabatically connected to a spin-liquid state with emergent gauge
excitations. Such spin liquids can be classified by the gauge
group determining the quantum numbers carried by the neutral
$S=1/2$ spinon excitations and the gauge excitations, and previous
work~\cite{z2,sf} has shown that the most prominent examples are
$Z_2$ and U(1) spin liquids. The $Z_2$ spin liquids are stable in
all spatial dimensions $d \geq 2$, while the U(1) spin liquids
exist only in $d \geq 3$ (the latter correspond to the existence
of a Coulomb phase in a compact U(1) gauge theory in $d\geq 3$, as
discussed in Ref.~\onlinecite{bosfrc}). Correspondingly, we also
have the metallic $Z_2$ FL$^*$ and U(1) FL$^*$ states. Our
previous work~\cite{ssv1} considered primarily the $Z_2$ FL$^*$
state, whereas here we focus on the U(1) FL$^*$ state.

As we have already discussed, these non-magnetic states may lead
to magnetic order at low energies, or in proximate states in a
generalized phase diagram. In this manner the FL state leads to
the SDW state, while the FL$^*$ states lead to the $Z_2$ SDW$^*$
and the U(1) SDW$^*$ states. The relation between the metallic SDW
and SDW$^*$ states has a parallel to that between the insulating
N\'{e}el state and the AF$^*$ state of Refs.~\onlinecite{bfn,sf}.

We will also discuss the evolution from the U(1) SDW$^*$ state to
the conventional Fermi liquid. As explained earlier, the
underlying transition is that between FL and FL$^*$ states which
controls the nature of the Fermi surface. In Ref.~\onlinecite{ssv1},
we argued that the spinon pairing in the $Z_2$ FL$^*$ state
implied that there must be a superconducting state in between the
FL and $Z_2$ FL$^*$ states. There is no such pairing in the U(1)
FL$^*$ state, and hence there is the possibility of a direct
transition between the FL and U(1) FL$^*$ states: this transition
and the nature of the states flanking it are the foci of our
paper. Note that the volume of the Fermi surface {\em jumps} at
this transition. Nevertheless the transition may be second order.
This is made possible by the vanishing of the quasiparticle
residue on an entire portion of the Fermi surface (a ``hot'' Fermi
surface) on approaching the transition from the FL side. Non-Fermi
liquid physics is clearly to be expected at such a second order
Fermi-volume changing transition. We reiterate that the U(1)
FL$^*$ state is only believed to exist in $d>2$.

\begin{figure}
\epsfxsize=3in \centerline{\epsffile{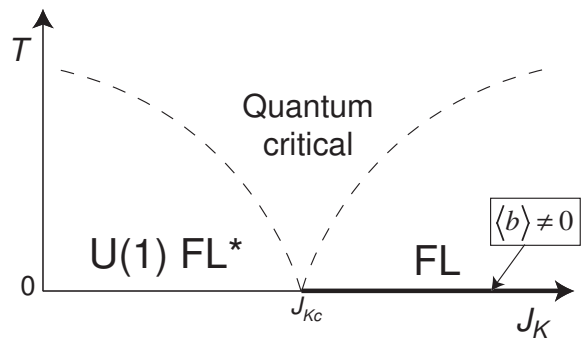}}
\caption{Crossover phase diagram for the vicinity of the $d=3$
quantum transition involving breakdown of Kondo screening. $J_K$
is the Kondo exchange in the Hamiltonian introduced in
Section~\protect\ref{mft}. The only true phase transition above is
that at the $T=0$ quantum critical point at $J_K = J_{Kc}$ between
the FL and FL$^*$ phases. The ``slave'' boson $b$ measures the
mixing between the local moments and the conduction electrons and
is also described in Section~\protect\ref{mft}. The crossovers are
similar to those of a dilute Bose gas as a function of chemical
potential and temperature, as discussed in
Refs.~\protect\onlinecite{weichmann,book}---the horizontal axis is
a measure of the boson chemical potential $\mu_b$. The boson is
coupled to a compact U(1) gauge field; at $T=0$ this gauge field
is in the Higgs/confining phase in the FL state, and in the
deconfining/Coulomb phase in the FL$^*$ state. There is no phase
transition at $T>0$ between a phase with $\langle b \rangle \neq
0$ and a phase with $\langle b \rangle =0$ because such a
transition is absent in a theory with a {\em compact} U(1) gauge
field in $d=3$ \cite{compact} (the mean-field theories of
Sections.~\protect\ref{mft} and \protect\ref{mfmag} do show such
transitions, but these will turn into crossovers upon including
gauge fluctuations). The compactness of the gauge field therefore
plays a role in the crossovers in the ``renormalized classical''
regime above the FL state (this has not been worked out in any
detail here). However, the compactness is not expected to be
crucial in the quantum-critical regime. The crossover line
displayed between the FL and quantum critical regimes can be
associated with the ``coherence'' temperature of the heavy Fermi
liquid. At low $T$, as discussed in the text, there are likely to
be additional phases associated with magnetic order (the SDW and
SDW$^*$ phases), and these are not shown above but are shown in
Fig.~\protect\ref{qcusdw}; they also appear in the mean-field
phase diagram in Fig.~\protect\ref{pht1}. } \label{cross}
\end{figure}

We study the FL and U(1) FL$^*$ states by the ``slave'' boson
method, introduced in the context of the single-moment Kondo
problem~\cite{hewson}. In this method, the condensation of the
slave boson marks the onset of Kondo coherence that characterizes
the FL phase. In contrast the slave boson is not condensed in the
FL$^*$ phase. Fluctuations about this mean-field description lead
to the critical theory of the transition involving a propagating
boson coupled to a compact U(1) gauge field, in the presence of
damping from fermionic excitations.

We note that earlier studies~\cite{newns,coleman} of
single-impurity problems found a temperature-induced mean-field
transition between a state in which the slave boson is condensed
(and hence the local moment is Kondo screened) and a state in
which the boson has no condensate: however, it was correctly
argued that this transition is an artifact of the mean-field
theory, and no sharp transition exists in the single-moment Kondo
problem at $T>0$.
If we now naively generalize this single-impurity model to the lattice,
we will find that the $T=0$ ground state always has Kondo screening.
It is only upon including frustrating inter-moment exchange
interactions -- equivalent to having ``dispersing'' spinons --
that it is possible to breakdown Kondo screening and reach a state in
which the slave boson is not condensed.
This transition is not an artifact of mean-field theory, we
show here that it remains sharply defined in $d=3$.

Our analysis of the above $d=3$ U(1) gauge theory leads to the
schematic crossover phase diagram as a function of the Kondo
exchange $J_K$ and $T$ shown in Fig~\ref{cross}.

\begin{figure}
\epsfxsize=3in \centerline{\epsffile{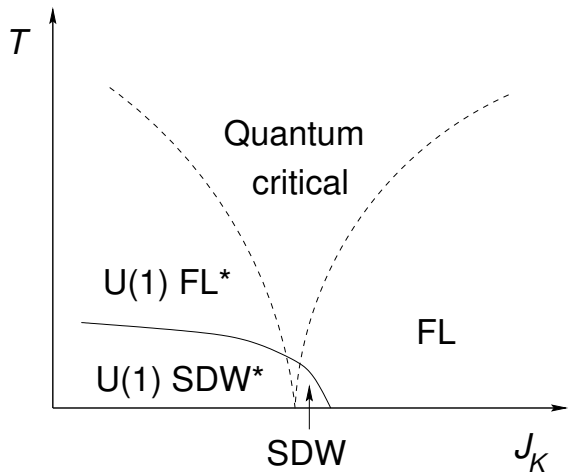}}
\caption{
Expected phase diagram and crossovers for the evolution from the
U(1) SDW$^*$  phase to the conventional FL.
Two different
transitions are {\em generically} possible at zero temperature: Upon moving from
the SDW$^*$ towards the Fermi liquid,
the fractionalization is lost first followed by the disappearance of magnetic
order. Nevertheless the higher temperature
behavior in the region marked `quantum critical' is non-fermi liquid like, and
controlled
by the Fermi volume changing transition from FL to FL$^*$. This may be loosely
associated to
the breakdown of Kondo screening.
}
\label{qcusdw}
\end{figure}

The crossover phase diagram in Fig. \ref{cross} is similar
to that of a dilute Bose
gas as a function of chemical potential and temperature
\cite{weichmann,book}.
Here the bosons are coupled to a U(1) gauge field, and this is important
for many of the critical properties to be described in the body of the
paper.
Notably, in Fig.~\ref{cross} the density of bosons is {\em not}
fixed, and varies as a function of $T$, $J_K$ and other couplings
in the Hamiltonian. Indeed, the contours of constant boson density
have a complicated structure, which are similar to those in
Ref.~\onlinecite{book}. This variation in the boson density is a
crucial distinction from earlier analyses~\cite{ki,nagaosa} of
boson models coupled to damped U(1) gauge fields: in these earlier
works, the boson density was fixed at a $T$-independent value. As
we will see, allowing the boson density to vary changes the
critical properties, and has significant consequences for the
structure of the crossover phase diagram and for the $T$
dependence of observables.

We will show that non-Fermi liquid physics obtains in the quantum
critical region of this transition. Furthermore, we argue that
fluctuation effects may lead to a spin density wave developing out
of the spinon Fermi surface of the U(1) FL$^*$ phase, thereby
obtaining the U(1) SDW$^*$ phase. The expected phase diagram and crossovers
for the evolution from the U(1) SDW$^*$ phase to the FL phase is shown in Fig.
\ref{qcusdw}. We
examine few
 different kinds
of such U(1) SDW$^*$ phases depending on the details of the spinon
Fermi surface. We also describe a number of specific experimental
signatures of the U(1) SDW$^*$ phase which may help to distinguish
it from more conventional magnetic metals.

\subsection{Relation to earlier work}
\label{earlier}

We have already mentioned a number of precursors to our ideas in
our discussion so far. Here, we complete this by noting some other
related developments in the literature.

Early on, Andrei and Coleman \cite{piers2} and Kagan {\em et al.}
\cite{kagan} discussed the possibility of the decoupling of local
moments and conduction electrons in Kondo lattice models. Andrei
and Coleman had the local moments in a spin-liquid state which is
unstable to U(1) gauge fluctuations, and did not notice violation
of Luttinger's theorem. The possibility of small electronic Fermi
surfaces was noted by Kagan {\em et al.}, but no connection was
made to the requirement this imposes on emergent gauge excitations
\cite{ssv1}.

More recently, Burdin {\em et al.} described many aspects of the
physics we are interested in a dynamical mean-field theory of a
random Kondo lattice~\cite{bgg}. In this work, they obtained a
state in which local moments formed a spin liquid and stayed
essentially decoupled from the conduction electrons. They
emphasized that the transition between such a state (which is the
analog of our FL$^\ast$ states) and a conventional heavy Fermi
liquid (the FL state) should be understood as a Fermi volume
changing transition. However questions of emergent gauge structure
were not addressed by them.

Demler {\em et al.} \cite{demler} discussed fractionalized phases
of Kondo lattice models. However, they did not consider any states with
long-lived electron-like quasiparticles, as are present in the
FL$^*$ phase.

Recently Essler and Tsvelik \cite{tsvelik} discussed the fate of
one-dimensional Mott insulator under a particular long-range
inter-chain hopping. At intermediate temperatures, they obtain a
state with a small Fermi surface, in that the Fermi surface volume
does not count the local moments \cite{foot1}. However, their
construction does not lead to a state with emergent gauge
excitations in higher dimensions, and as they conclude, their
state is unstable to magnetic order at low temperatures. We
believe this low-$T$ state is an ordinary SDW state, and any
realizations of small Fermi surfaces at intermediate temperatures
are remnants of one-dimensional physics. In contrast, all our
constructions are genuinely higher-dimensional, and only work for
$d \geq 2$.

The physics of the destruction of the large Fermi surface by the
vanishing of Kondo screening has been addressed in interesting
recent works~\cite{si,sun,zhu} using an ``extended dynamical
mean-field theory''. We have argued in our discussion above that
vanishing of Kondo screening is conceptually quite a different
transition from the onset of magnetic order; consistent with this
expectation, Sun and Kotliar \cite{sun} found two distinct points
associated with these transitions.
It is our contention that the critical theory
of the FL to U(1) FL$^*$ transition (discussed in the present
paper) is the $d=3$ realization of the large-dimensional critical
point with vanishing Kondo screening found by Sun and Kotliar.

\subsection{Outline}
\label{outline}

The rest of the paper is organized as follows: In Section
\ref{flstar}, we briefly review the properties of various
fractionalized Fermi liquids (FL$^*$). A specific U(1) FL$^*$
state where the spinons form a Fermi surface is considered. In
Section \ref{mft}, we construct a mean-field description of this
state and its transition to the heavy Fermi liquid. This
transition involves a jump in the Fermi surface volume but is
nevertheless shown to be second order within the mean-field
theory. This is made possible by the vanishing of the
quasiparticle residue $Z$ on an entire Fermi surface (a ``hot''
Fermi surface) as one moves from the heavy Fermi liquid to the
fractionalized Fermi liquid. Fluctuations about this mean-field
description are then considered. In Section \ref{fluct}, we first
consider fluctuation effects on the phases -- in particular the
FL$^*$ phase. We argue that the specific heat coefficient $\gamma$
diverges logarithmically once the leading-order fluctuations are
included. Furthermore, fluctuations also make possible a spin
density wave instability of the spinon Fermi surface, leading to a
U(1) SDW$^*$ state. To illustrate possible phases, we will discuss
an improved mean-field theory which includes the SDW order
parameter, and present phase diagrams showing the influence of
temperature and magnetic field.
We then examine fluctuation
effects at the critical point of the transition between FL$^*$ and
FL in Section \ref{flcrt}.
We argue that the logarithmic divergence of the specific heat
coefficient persists in the quantum critical region, and also
that non-Fermi liquid transport obtains there.
In Section \ref{sdwstar}, we
discuss the properties of the U(1) SDW$^*$ phase in greater detail
with particular attention to its identification in experiments.
A discussion of the implications for various experiments in
Sec.~\ref{disc} will conclude the paper.

%%%%%%%%%%%%%%%%%%%%%%%%%%%%%%%%%%%%%%%%%%%%%%%%%%%%%%%%%%%%%%%%%%

\section{Fractionalized Fermi liquids}
\label{flstar}

The existence of non-magnetic translation invariant ``small Fermi
surface'' states was shown in a recent article by us \cite{ssv1}
with a focus on two-dimensional Kondo lattices. Such states were
obtained when the local-moment system settles into a
fractionalized spin liquid (rather than a magnetically ordered
state) due to inter-moment interactions. A weak Kondo coupling to
conduction electrons does not disrupt the structure of the spin
liquid but leaves a sharp Fermi surface of quasiparticles whose
volume counts the conduction electron density alone (a small Fermi
surface).
Thus these states have fractionalized excitations that coexist
with conventional Fermi-liquid-like quasiparticle excitations. We
dubbed these states FL$^*$ (to distinguish them from the
conventional Fermi liquid FL). We also pointed out an intimate
connection between the disappearance of the large Fermi surface
and fractionalization, and this is discussed further in
Appendix~\ref{oshi}.

The FL$^*$ phase can be further classified by the nature of the
spin liquid formed by the local moments. Recent years have seen
considerable progress in our understanding of fractionalized spin
liquids. An important feature of spin liquid states in $d \geq 2$
is that they possess emergent gauge structure. Put simply, this
means that the distinct excitations in such phases interact with
each other through long ranged interactions which can be
mathematically encapsulated as gauge interactions. In other words,
the effective field theory of the state is a gauge theory in its
deconfined phase. The two natural possibilities are that the
emergent gauge group is either $Z_2$ or U(1). The former is
allowed in any dimension $d \geq 2$ while the latter is only
allowed in $d = 3$ (or higher).

The $Z_2$ states have been discussed at length in the literature
and in the present context in our earlier work \cite{ssv1}. In
contrast, the U(1) states have not been discussed much, though
their possible occurrence (in $d = 3$) and their universal
properties have been appreciated by many workers in the field. We
therefore provide a quick discussion: The distinct excitations in
the $d = 3$ U(1) spin liquid phases are neutral spin-$1/2$
spinons, a gapless (emergent) gauge photon, and a gapped point
defect (the ``monopole''). The spinons are minimally coupled to
the photon and hence interact through emergent long ranged
interactions. For simple microscopic models that realize such
phases, see Ref.~\onlinecite{bosfrc,hermele}.
A crucial distinction between the $Z_2$ spin liquids is that
the spinons in this phase are {\em not} generically paired,
{\em i.e.}, the spinon number is conserved~\cite{foot2}.

Several classes of spin liquids are theoretically possible with
the same gauge structure. These may be characterized by the
statistics of the spinons, their band structure, etc. For the rest
of this paper, we will focus on a particular three-dimensional
U(1) spin liquid state with fermionic spinons that form a Fermi
surface. A specific toy model which displays this phase is
presented in Appendix~\ref{toy}.

As with the $Z_2$ spin liquids discussed in
Ref.~\onlinecite{ssv1}, the gauge structure in the U(1) spin
liquid state is also stable to a weak Kondo coupling to conduction
electrons~\cite{foot3}. The resulting U(1) FL$^*$ state consists
of a spinon Fermi surface coexisting with a separate Fermi surface
of conduction electrons. There will also be gapless photon and
gapped monopole excitations. The physical electron Fermi surface
(as measured by de-Haas van Alphen experiments for instance) will
have a small volume that is determined by the conduction electrons
alone.

In our previous work, we pointed out that the transition from a
$Z_2$ FL$^*$ phase to the heavy FL will generically be preempted
by superconductivity. This is due to the pairing of spinons in the
$Z_2$ phase. In contrast, we expect that due to conservation of
spinon number a direct transition between the U(1) FL$^*$ and
heavy FL phases should be possible.

%%%%%%%%%%%%%%%%%%%%%%%%%%%%%%%%%%%%%%%%%%%%%%%%%%%%%%%%%%%%%%%%%%

\section{Mean-field theory}
\label{mft}

A simple mean-field theory allows a description both of a U(1)
FL$^*$ phase and its transition to the heavy FL.
Consider a three-dimensional Kondo-Heisenberg model, for
concreteness on a cubic lattice:
\begin{eqnarray}
  H &=& \sum_k \epsilon_k c^{\dagger}_{k\alpha} c_{k\alpha} +
  \frac{J_K}{2}\sum_r
  \vec S_r \cdot c^{\dagger}_{r\alpha} \vec \sigma_{\alpha\alpha'} c_{r\alpha'}
\nonumber\\
  &+&
  J_H \sum_{\langle rr' \rangle}\vec
  S_r \cdot \vec S_{r'} \,.
\label{KH}
\end{eqnarray}
Here $c_{k\alpha}$ represent the conduction electrons and $\vec
S_r$ the spin-$1/2$ local moments on the sites of a cubic lattice,
summation over repeated spin indices $\alpha$ is implicit. We use
a fermionic ``slave-particle'' representation of the local
moments:
\begin{equation}\label{sf}
\vec S_r = \frac{1}{2}
f^{\dagger}_{r\alpha} \vec \sigma_{\alpha\alpha'} f_{r\alpha'}
\end{equation}
where $f_{r\alpha}$ describes a spinful fermion destruction
operator at site $r$.

Proceeding as usual, we consider a decoupling of both the Kondo
and Heisenberg exchange using two auxiliary fields in the
particle-hole channel. Treating the fluctuations of these
auxiliary fields by a saddle point approximation (formally
justified for a large-$N$ SU($N$) generalization), we obtain the
mean-field Hamiltonian
\begin{eqnarray}\label{mf1}
  H_{\rm mf} & = & \sum_k \epsilon_k c^{\dagger}_{k\alpha} c_{k\alpha} -
  \chi_0\sum_{\langle rr' \rangle} \left(f^{\dagger}_{r\alpha} f_{r'\alpha} +
\mbox{
  h.c.}\right)
    \nonumber\\
  &
  + & \mu_f\sum_r f^{\dagger}_{r\alpha}f_{r\alpha}- b_0 \sum_k
\left(c^{\dagger}_{k\alpha}
  f_{k\alpha} + \mbox{h.c.} \right)
\end{eqnarray}
where we assumed $\chi_0$ and $b$ to be real, and have dropped additional
constants to $H$.
The mean-field parameters $b_0, \chi_0, \mu_f$ are determined by
the conditions
\begin{eqnarray}
1 &=& \langle f^{\dagger}_{r\alpha} f_{r\alpha} \rangle  \,,\\
2 b_0 & = & J_K \langle c^{\dagger}_{r\alpha} f_{r\alpha} \rangle\,, \\
2 \chi_0 & = & J_H \langle f^{\dagger}_{r\alpha} f_{r'\alpha} \rangle \,.
\end{eqnarray}
In the last equation $r,r'$ are nearest neighbors.

%\subsection{Mean-field transition between FL and FL$^\ast$}

%The mean-field Hamiltonian is readily diagonalized.
There are two
qualitatively different zero-temperature phases. First, there is
the usual Fermi liquid (FL) phase when $b_0, \chi_0, \mu_f$ are
all non-zero. (Note that $b_0 \neq 0$ implies that $\chi_0 \neq
0$). This phase is readily seen to have a large Fermi surface as
expected. Second, there is a phase (FL$^*$) where the Kondo
hybridization $b_0 = 0$ but $\chi_0 \neq 0$. (In this phase $\mu_f
= 0$.) This mean-field state represents a situation where the
conduction electrons are decoupled from the local moments and form
a {\em small} Fermi surface. The local-moment system is described
as a spin fluid with a Fermi surface of neutral spinons. We expect
that $\chi_0 \sim J_H$.

The transition between these two different states can also be
examined within the mean-field theory. Interestingly, the
transition is second order (despite the jump in Fermi volume) and
is described by $b_0 \rightarrow 0$ on approaching it from the
Fermi liquid side. How can a second order transition be associated
with a jump in the volume of the electron Fermi surface? This can
be understood by examining the Fermi surfaces closely in this
mean-field theory.

The mean-field Hamiltonian is diagonalized by the transformation
\begin{eqnarray}
c_{k\alpha} & = & u_k \gamma_{k\alpha +} + v_k \gamma_{k\alpha -},
\nonumber \\
f_{k\alpha} & = & v_k \gamma_{k\alpha+} - u_k \gamma_{k\alpha-}.
\end{eqnarray}
Here $\gamma_{k\alpha\pm}$ are new fermionic operators in terms of
which the Hamiltonian takes the form
\begin{equation}\label{hdiag}
  H_{\rm mf} = \sum_{k\alpha}E_{k+}\gamma_{k\alpha +}^{\dagger}\gamma_{k\alpha
  +} + E_{k-}\gamma_{k\alpha -}^{\dagger}\gamma_{k\alpha -},
\end{equation}
with
\begin{equation}\label{energy}
  E_{k\pm} = \frac{\epsilon_k + \epsilon_{kf}}{2} \pm
  \sqrt{\left(\frac{\epsilon_k - \epsilon_{kf}}{2}\right)^2 + b_0^2}.
\end{equation}
Here $\epsilon_{kf} = \mu_f - \chi_0\sum_{a = 1,2,3} \cos(k_a)$.
The $u_k, v_k$ introduced above are determined by
\begin{eqnarray}
\label{ukvk}
u_k  =  -\frac{b_0v_k}{E_{k+} - \epsilon_k} \,,~~
u_k^2 + v_k^2  =  1 \,.
\end{eqnarray}

Consider first the FL$^*$ phase where $b_0 = 0 = \mu_f$, but
$\chi_0 \neq 0$. The electron Fermi surface is determined by the
conduction electron dispersion $\epsilon_k$ and is small. The
spinon Fermi surface encloses one spinon per site and has volume
half that of the Brillouin zone. For concreteness, we will
consider the situation where the electron Fermi surface does not
intersect the spinon Fermi surface. We will also assume that the
conduction electron filling is less than half.

\begin{figure}[t]
\epsfxsize=3.3in
\centerline{\epsffile{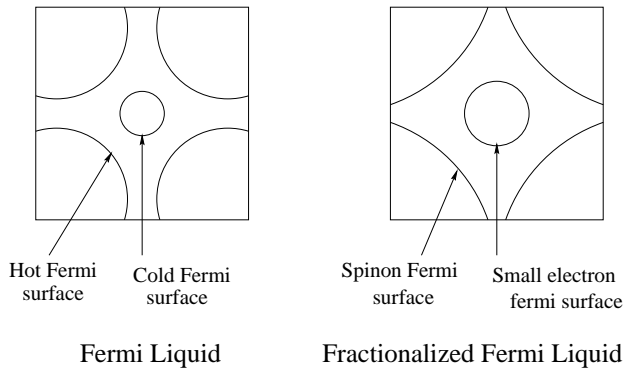}} \caption{ Fermi
surface evolution from FL to FL$^*$: close to the transition, the
FL phase features two Fermi surface sheets (the cold $c$ and the
hot $f$ sheet, see text). Upon approaching the transition, the
quasiparticle residue $Z$ on the hot $f$ sheet vanishes. On the
FL$^*$ side, the $f$ sheet becomes the spinon Fermi surface,
whereas the $c$ sheet is simply the small conduction electron
Fermi surface. } \label{fig:hotfs}
\end{figure}

Now consider the FL phase near the transition (small $b_0$). In
this case, there are two bands corresponding to $E_{k\pm}$: one
derives from the $c$-electrons (with weak $f$ character) while the
other derives from the $f$-particles (with weak $c$-character). We
will call the former the $c$-band and the latter the $f$-band.
For small $b_0$, both bands intersect the Fermi energy so that the
Fermi surface consists of two sheets (see Fig.~\ref{fig:hotfs}).
The total volume is large, {\em i.e}, includes both local moments and
conduction electrons. Upon moving toward the transition to FL$^*$
($b_0$ decreasing to zero), the $c$-Fermi surface expands in size
to match onto the small Fermi surface of FL$^*$. On the other
hand, the $f$-Fermi surface shrinks to match onto the {\em spinon}
Fermi surface of FL$^*$.

Upon increasing $b_0$ in the FL state and depending on the band
structure, another transition is possible, where the $c$ band
becomes completely empty.
Then, the Fermi surface topology changes from two sheets to
a single sheet -- such a transition between two conventional
Fermi liquids is known as Lifshitz transition and will not
be further considered here.

The quasiparticle weight $Z$ close to the FL--FL$^*$ transition
is readily calculated in the present mean-field theory.
For the electron Green's function we find
\begin{equation}\label{gf}
  {\cal G}(k, i\omega_{\nu}) = \frac{u_k^2}{i\omega_{\nu} -
  E_{k+}} + \frac{v_k^2}{i\omega_{\nu} - E_{k-}}.
\end{equation}
Therefore at the Fermi surface of the $c$-band (which has
dispersion $E_{k+}$, the quasiparticle residue $Z = u_k^2$.
At
this Fermi surface, $E_{k+} \approx \epsilon_{k} \approx 0$ so that
\begin{equation}\label{Ekplus}
  E_{k+} \approx \epsilon_k + \frac{b_0^2}{\epsilon_k -
  \epsilon_{kf}} \Rightarrow u_k \approx -\frac{J_H}{b_0}v_k.
\end{equation}
Using Eqs. (\ref{ukvk}), we then find $Z \approx 1$ on the
$c$-Fermi surface.

At the Fermi surface of the $f$-band on the other hand, $Z =
v_k^2$. Also near this Fermi surface, $|\epsilon{k} - \epsilon_{kf}| \approx t$
where $t$ is the conduction
electron bandwidth. We have assumed as is reasonable that
$t \gg J_H$.
Thus for the $f$-Fermi surface,
\begin{equation}\label{Ekhot}
  E_{k+} \approx \epsilon_k + \frac{b_0^2}{\epsilon_k -
  \epsilon_{kf}} \Rightarrow u_k \approx -\frac{t}{b_0}v_k.
\end{equation}
This then gives
\begin{equation}\label{Zhot}
  Z = v_k^2 \approx \left(\frac{b_0}{t}\right)^2.
\end{equation}
Thus the quasiparticle residue stays non-zero on the $c$-Fermi
surface while it decreases continuously to zero on the $f$-Fermi
surface on moving from FL to FL$^*$.
(The $f$-Fermi surface is ``hot'' while the $c$-Fermi surface
is ``cold''.)

Clearly the critical point is not a Fermi liquid. $Z$ vanishes
throughout the hot Fermi surface at the transition, and non-Fermi
liquid behavior results. It is interesting to contrast this result
with the spin-fluctuation model (Hertz-Moriya-Millis criticality)
where the non-Fermi liquid behavior is only associated with some
``hot'' lines in the Fermi surface, and consequently plays a
subdominant role.

Despite the vanishing quasiparticle weight $Z$, the effective mass $m^*$
of the large Fermi surface state does not diverge at the
transition in this mean-field calculation, because the electron
self-energy is momentum-dependent. Physically, the quasiparticle
at the hot Fermi surface is essentially made up of the
$f$-particle for small $b$; even when $b$ goes to zero the
$f$-particle (the spinon) continues to disperse due to the
non-vanishing $\chi_0$ term. Indeed the low-temperature specific
heat $C \sim \gamma T$ with $\gamma$ non-zero in both phases. As
we argue below, this is an artifact of the mean-field
approximation and will be modified by fluctuations.

The detailed shape of the spinon Fermi surface in the FL$^*$ phase
(or the hot Fermi surface which derives from it in the FL phase)
depends on the details of the lattice and the form of the local
moment interactions. For the particular model discussed above, the
spinon Fermi surface is perfectly nested. In more general
situations, a non-nested spinon Fermi surface will obtain. In all
cases, however, the volume of the spinon Fermi surface will
correspond to one spinon per site.

%%%%%%%%%%%%%%%%%%%%%%%%%%%%%%%%%%%%%%%%%%%%%%%%%%%%%%%%%%%%%%%%%%

\section{Fluctuations: Magnetism and singular specific heat}
\label{fluct}

Fluctuation effects modify the picture obtained in the mean-field
theory in several important ways. We first discuss fluctuation
effects in the two phases. The heavy Fermi liquid phase is of
course stable to fluctuations - their main effect being to endow
the $f$-particle with a physical electric charge thereby making it
an electron~\cite{read,millee}. Fluctuation effects are more interesting in the
FL$^*$ state, and are described by a U(1) gauge theory minimally
coupled to the spinon Fermi surface (which continues to be
essentially decoupled from the conduction electron small Fermi
surface). This may be made explicit by parameterizing the
fluctuations in the action in the FL$^*$ phase as follows:
\begin{equation}
\chi_{rr'}(\tau)  =  e^{ia_{rr'}(\tau)}\chi_{0rr'} \,.
\end{equation}
The action then becomes
\begin{eqnarray}
\label{sbact}
S & = & S_c + S_f + S_{fc} + S_b \,,\\
S_c & = & \int {\rm d}\tau \sum_k \bar{c}_k (\partial_\tau -
\epsilon_k)c_k \nonumber \,,\\
S_f & = & \int {\rm d}\tau \sum_r \bar{f}_r(\partial_\tau - ia_0)f_r \nonumber \\
& &  -\sum_{\langle rr'\rangle}\chi_0
\left(e^{ia_{rr'}}\bar{f}_rf_{r'} + \mbox{h.c.} \right) \nonumber\,, \\
S_{cf} & = & -\int {\rm d}\tau \sum_r \left(b_r\bar{c}_rf_r +
\mbox{h.c.} \right) \nonumber\,, \\
S_b & = & \int {\rm d}\tau \sum_r \frac{4|b_r|^2}{J_K} \,. \nonumber
\end{eqnarray}
As usual, the field $a_0$ is introduced to impose the constraint
that there is one spinon per site and may be interpreted as the
time component of the gauge field.
By assumption $b_r$ is not condensed.
It is useful to start by completely ignoring all
coupling between $c$ and $f$ fermions. The action for the $f$
particles describes a Fermi surface of spinons coupled to a
compact U(1) gauge field.

An important simplification for the three-dimensional systems of
interest (as compared to $d = 2$) is that the U(1) gauge theory
admits a deconfined phase where the spinons potentially survive as
good excitations of the phase. In what follows we will assume that
the system is in such a deconfined phase. (This is formally
justified in the same large-$N$ limit as the one for the mean
field approximation.)
This deconfined phase has  a Fermi surface
of spinons coupled minimally to a gapless ``photon'' (U(1) gauge
field). (Due to the compactness of the underlying gauge theory,
there is also a gapped monopole excitation.) Thus two static
spinons interact with each other through an {\em emergent long
range} $1/r$ Coulomb interaction. Putting back a small coupling
between the $c$ and $f$ particles will not change the deconfined
nature of this phase. (In particular the monopole gap will be
preserved.)
This is the advocated U(1) FL$^*$ phase.

\subsection{Specific heat}
\label{sec:specific}

The coupling of the massless gauge photon to the spinon Fermi
surface leads to several interesting modifications of the mean
field results.  First, consider the effect of the spatial
components of the gauge field. It is useful to work in the gauge
$\vec \nabla \cdot \vec a = 0$ so that the vector potential is
purely transverse. Unless otherwise stated, we assume a generic
spinon Fermi surface (without flat portions) henceforth.
Integrating out the spinons and expanding the resulting action to
quadratic order gives the following well-known form for the
propagator for these transverse gauge fluctuations:
\begin{eqnarray}
D_{ij}(\vec k, i\omega_n) &\equiv& \langle a_i(\vec k,
i\omega_n)a_j(-\vec k, -i\omega_n) \rangle \nonumber \\ &=&
\frac{\delta_{ij} - k_ik_j/k^2}{\Gamma |\omega_n|/k + \chi_f k^2} \,.
\label{dprop}
\end{eqnarray}
Here $\Gamma, \chi_f$ are positive constants that are determined
by the details of the spinon dispersion, and $\omega_n$ is an
imaginary Matsubara frequency. Note that the gauge fluctuations
are overdamped in the small $q$ limit. As was first shown in a
different context by Holstein {\em et al.}~\cite{holst} (and
reviewed in Appendix~\ref{heatapp}), this form of the gauge field
action leads to a $T\ln 1/T$ singularity in the low-temperature
specific heat. Thus the specific heat coefficient $\gamma = C/T$
diverges logarithmically at low temperature in the U(1) FL$^*$
phase.

We also briefly mention the effect of the longitudinal
(time-component) of the gauge field. This couples to the local $f$
fermion density, and so its influence is very much like a
repulsive density-density interaction. The longitudinal gauge
field propagator has a structure very similar to that of a
standard RPA density fluctuation propagator, and so does not lead
to any non-Fermi liquid behavior.

\subsection{Magnetic instability}
\label{sec:magins}

The repulsive interaction mediated by the longitudinal part of the
gauge interaction can lead to various instabilities of the spinon
Fermi surface.  In particular, it is interesting to consider an
SDW instability of the spinon Fermi surface. The resulting state
will have magnetic long range order that could potentially have a
weak moment as it is an SDW state that is formed out of the spinon
Fermi surface. However, in contrast to the traditional view of the
weak magnetism, here the SDW instability is {\em not} that of the
large Fermi surface heavy Fermi liquid. Despite the occurrence of
magnetic long range order, this magnetic state is far from
conventional. Because the SDW order parameter is gauge neutral,
the presence or absence of a SDW condensate has little substantive
effect on the structure of the gauge fluctuations. Indeed, the
latter remain as in the U(1) FL$^{\ast}$ state, even after the
magnetic order has appeared in the descendant U(1) SDW$^*$ state.
The spinons continue to be deconfined and are coupled to a gapless
U(1) gauge field. Further, the monopole survives as a gapped
excitation -- this yields a sharp distinction with more
conventional magnetic phases. These gauge excitations coexist with
the gapless magnons associated with broken spin rotation
invariance and with a Fermi surface of the conduction electrons.
However, due to the broken translational symmetry in this state,
there is no sharp distinction between small and large Fermi
surfaces. So to reiterate, the exotic magnetic metal, dubbed U(1)
SDW$^*$, emerges as a low-energy instability of the spinon Fermi
surface of the parent U(1) FL$^*$ state.

Different possibilities emerge for the formation of the spin
density wave out of the parent U(1) FL$^*$ phase, depending on the
details of the spinon Fermi surface and the strength of the
interactions driving the SDW instability. We enumerate some of
them below:
\begin{itemize}
\item
(A) Perfectly nested spinon Fermi surface:

In this case, arbitrarily weak interactions will drive an SDW
instability. In the resulting state, the spinons are gapped. So
upon integrating out the spinons, the effective action for the
gauge field can be expanded safely in spatial and temporal
gradients, with no long-range couplings. Gauge invariance now
demands that these terms in the gauge field action have the
standard Maxwell form. Consequently, the photon becomes a sharp
propagating mode at low energies (below the spinon gap) with
linear dispersion. Despite clearly being a distinct phase from
conventional spin density wave metals, the experimental
distinction is subtle.

\item (B) Generic spinon Fermi surface, weak interaction:

For a generic spinon Fermi surface, the leading spin density wave
instability (which will require an interaction strength beyond
some threshold value) will be at a wavevector that matches one of
the ``$2k_F$'' wavevectors of the spinon Fermi surface. In the
resulting state, a portion of the spinon Fermi surface (away from
points connected by the ordering wavevector) survives intact. The
damping of the gapless U(1) gauge fluctuations due to coupling to
gapless spinons is preserved. Consequently the low-temperature
specific heat will continue to behave as $C(T) \sim T\ln (1/T)$.
Thus for this particular U(1) SDW$^*$ state its non-Fermi liquid
nature is readily manifested by specific heat measurements,
providing a concrete example of a weak moment SDW metal with
non-Fermi liquid thermodynamics at low temperature.

\item
(C) Generic spinon Fermi surface, strong interaction:

If the interactions are strong enough, even for a non-nested
spinon Fermi surface, the spinons can develop a full gap with no
portion of their Fermi surface remaining intact. The resulting
phase is the same as that obtained in (A), and has a sharp
propagating linear dispersing photon at low energies.

\end{itemize}

In Section~\ref{sdwstar} we discuss experimental probes that can
help distinguish these U(1) SDW$^*$ phase from the conventional
spin density wave metals.

\subsection{Mean-field theory with magnetism}
\label{mfmag}

In view of the possible occurrence of SDW phases we will now
consider a modified mean-field theory which captures the magnetic
instability at the mean-field level, but does no longer correspond
to a large-$N$ saddle point. We will discuss the fully
self-consistent solution of the mean-field equations for arbitrary
temperature and external magnetic field.

\begin{figure}[t]
\epsfxsize=3.1in \centerline{\epsffile{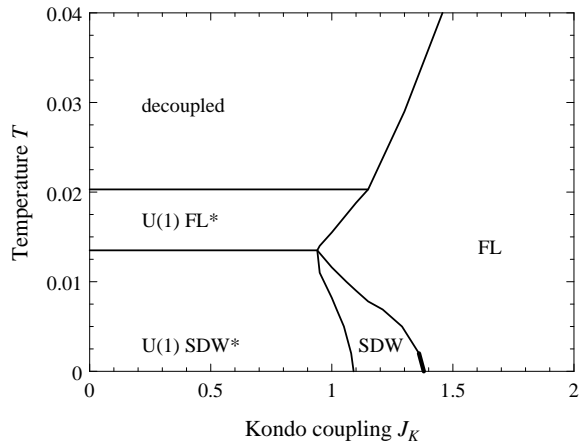}} \caption{
Mean-field phase diagram of $H_{\rm mf}$ (\protect{\ref{mf2}}) on
the cubic lattice, as function of Kondo coupling $J_K$ and
temperature $T$. Parameter values are electron hopping $t=1$,
Heisenberg interaction $J_H=0.1$, decoupling parameter $x=0.2$,
and conduction band filling $n_c=0.7$. Thin (thick) lines are
second (first) order transitions. The ``decoupled'' phase is an
artifact of the mean-field theory, and the corresponding
transitions will become crossovers upon including fluctuations, as
will the transition between the FL and U(1) FL$^*$ phases; the
transitions surrounding the SDW and SDW$^*$ phases will of course
survive beyond mean-field theory. } \label{pht1}
\end{figure}

\begin{figure}[t]
\epsfxsize=3.1in \centerline{\epsffile{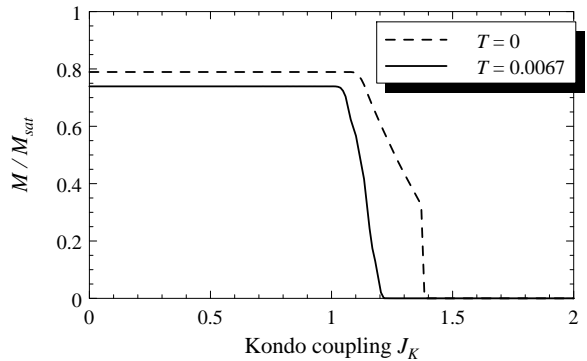}}
\caption{
Staggered magnetization determined from the mean-field solution
$H_{\rm mf}$ (\protect{\ref{mf2}}).
Parameter are as in Fig.~\protect\ref{pht1},
the two curves correspond to two horizontal cuts of the phase diagram
in Fig.~\protect\ref{pht1}. At $T=0$, the first-order character of the
SDW--FL transition is clearly seen.
Note that smaller values of the decoupling parameter $x$ yield smaller
values of the magnetization in the SDW and SDW$^*$ phases.
}
\label{mag1}
\end{figure}

The mean-field Hamiltonian, written down explicitly for SU(2)
symmetry, takes the following form:
\begin{eqnarray}
  H_{\rm mf} & = & \sum_{k} \epsilon_k c^{\dagger}_{k\alpha} c_{k\alpha} -
  \sum_{\langle rr'\rangle} \left(\chi_{rr'}^* f^{\dagger}_{r\alpha}
f_{r'\alpha} + \mbox{h.c.}\right)
    \nonumber\\
  &
  + & \sum_{r} \mu_{f,r} f^{\dagger}_{r\alpha}f_{r\alpha}
  -  \sum_{r} b_r \left(c^{\dagger}_{r\alpha} f_{r\alpha} + \mbox{h.c.} \right)
  \nonumber\\
  &+& \frac{1}{2} \sum_r (\vec{H}_{{\rm eff},r} + \vec{H}_{\rm ext})
  \cdot f^{\dagger}_{r\alpha} \vec \sigma_{\alpha\alpha'}
  f_{r\alpha'} \nonumber \\
  &+& \frac{1}{2} \vec{H}_{\rm ext} \cdot \sum_r
  c^{\dagger}_{r\alpha} \vec \sigma_{\alpha\alpha'} c_{r\alpha'} + E_{\rm const}
\label{mf2}
\end{eqnarray}
where $\vec{H}_{\rm ext}$ is the external field, and we have
allowed for a spatial dependence of the mean-field parameters
$\mu_{f,r}$, $\chi_{rr'}$, $b_r$, $\vec{H}_{{\rm eff},r}$. They
have to be determined from the following equations:
\begin{eqnarray}
1 &=& \langle f^{\dagger}_{r\alpha}f_{r\alpha} \rangle \,, \\
2 b_r & = & J_K \langle c^{\dagger}_{r\alpha} f_{r\alpha} \rangle \,, \\
2 \chi_{rr'} & = & (1-x) J_H \langle f^{\dagger}_{r\alpha}f_{r'\alpha} \rangle
\,, \\
\vec{H}_{{\rm eff},r} & = & x J_H \sum_{r'} \vec{M}_{r'} \,,~
\vec{M}_r = \frac{1}{2} \langle f^{\dagger}_{r\alpha} \vec
\sigma_{\alpha\alpha'} f_{r\alpha'} \rangle \,,
\end{eqnarray}
where the last sum runs over the nearest neighbors $r'$ of site
$r$. We have introduced a parameter $x$ which allows to control
the balance between ordered local-moment magnetism and spin-liquid
behavior of the $f$ electrons. A value $x=1/2$ would correspond to
an unrestricted Hartree-Fock treatment of the original Heisenberg
interaction; we will employ values $x<1/2$ in order to model a
{\em weak} magnetic instability of the spinon Fermi surface state.
The constant piece of the Hamiltonian reads
\begin{eqnarray}
E_{\rm const} &=& - \sum_r \mu_{f,r} + \sum_{r} \frac{2
b_r^2}{J_K}
\\
&+& \sum_{rr'} \frac{2 |\chi_{rr'}|^2}{(1-x) J_H} - \frac{1}{2}
\sum_{r} \vec{H}_{{\rm eff},r} \cdot \vec{M}_r \nonumber \,.
\end{eqnarray}

For simplicity, we consider a simple cubic lattice, and assume a
tight-binding dispersion for the conduction electrons, $\epsilon_k
= - 2 t \sum_{a = 1,2,3} \cos(k_a) - \mu_c$, where $\mu_c$
controls the conduction band filling. The mean-field equations can
be self-consistently solved using a large unit cell, allowing for
spatially inhomogeneous phases~\cite{mfsaddle}. In this section we
restrict our attention to mean-field solutions where the
$\chi_{rr'} = \chi_0$ fields are real (time-reversal invariant)
and obey the full lattice symmetries, and $b_r = b_0$ is
site-independent. We employ a $2\times 1$ unit cell, then
antiferromagnetism is characterized by $\vec{M}_r \cdot \hat x =
M_s \exp(i {\bf Q} \cdot {\bf r})$ where ${\bf Q}=(\pi,\pi,\pi)$
is the antiferromagnetic wavevector, and $\hat x$ is the
magnetization axis (which is arbitrary in zero external field).

In Fig.~\ref{pht1} we show a phase diagram obtained from
self-consistently solving (\ref{mf2}) together with the above
mean-field equations at zero external magnetic field.
A U(1) FL$^*$ phase with $b_0=0$ and $\chi_0
\neq 0$ is realized at intermediate temperatures. As expected, it
is unstable to magnetic order at low $T$, resulting in a U(1)
SDW$^*$ ground state for small $J_K$ -- this phase has in addition
$M_s \neq 0$. For the present parameter values, the spinon Fermi
surface is gapped out in the SDW$^*$ phase. Increasing $J_K$
drives the system into the FL phase with $b_0 \neq 0$, $\chi_0
\neq 0$, and $M_s = 0$; at low temperatures a conventional SDW
phase intervenes where all $b_0$, $\chi_0$, $M_s$ are non-zero.
Note that the transition between FL and SDW is weakly first order
at low temperatures. At high temperature, the mean-field theory
only has a ``decoupled'' solution with $b_0=\chi_0=M_s=0$ -- this
decoupling is a well-known mean-field artifact and reflects the
presence of incoherent excitations.

In the FL phase, the above mentioned Lifshitz transition occurs at
$J_K \approx 1.7$ in the low-temperature limit, {\em i.e.}, for
$J_K > 1.7$ only a single Fermi surface sheet remains. Note that
this transition does not lead to strong singularities in the
mean-field parameters.

The staggered magnetization of the SDW and SDW$^*$ states as
determined from the mean-field solution are shown in
Fig.~\ref{mag1}; we can expect that fluctuation corrections will
significantly reduce these mean-field values. We have also studied
different values of the decoupling parameter $x$; in particular
smaller values of $x$ lead to a suppression of ordered magnetism
in favor of the non-magnetic FL$^*$ state, i.e., the SDW
instability of FL$^*$ is shifted to lower temperatures (and
becomes completely suppressed at small $x$); similarly, the
ordered moment in the SDW phases is decreased with decreasing $x$.

Interesting physics obtains when an external magnetic field is turned on,
and the corresponding mean-field phase
diagram is discussed in Appendix \ref{mfbext}.

%%%%%%%%%%%%%%%%%%%%%%%%%%%%%%%%%%%%%%%%%%%%%%%%%%%%%%%%%%%%%%%%%%

\section{Fluctuations near the Fermi volume changing transition}
\label{flcrt}

We now turn to the effects of fluctuations beyond the mean-field
theory at the phase transition between the FL and U(1) FL$^*$
phases. In mean-field theory, this transition occurs through the
condensation of the slave boson field $b$. Such a condensation
survives as a sharp transition beyond mean-field only when $T=0$.

We begin by observing that in the mean-field theory all the
important changes near the transition occur at the hot Fermi
surface. The cold Fermi surface (essentially made up of
$c$-particles) plays a spectator role. We therefore integrate out
the $c$-fields completely from the action in Eq. (\ref{sbact}) to
obtain an effective action involving the $b,f$ and gauge fields
alone. We also partially integrate out $f$ excitations well away
from the hot Fermi surface: this changes the $b$ effective action
from the simple local term in (\ref{sbact}), and endows it with
frequency and momentum dependence. In this manner we obtain the
following effective action at long distance and time scales:
\begin{eqnarray}
\label{sbcrt}
S & = & S_b + S_f \,, \\
S_b & = & \int {\rm d}\tau {\rm d}^3 r \Biggl[\bar{b}\left(\partial_{\tau} -
\mu_b - ia_0 - \frac{(\vec \nabla_r
- i\vec a)^2}{2m_b}\right)b  \nonumber \\
& &~~~~~~~~~~~~~~~~~~~~~~~~~  + \frac{u}{2}|b|^4 +.... \Biggr],
\end{eqnarray}
and $S_f$ has the same form as in (\ref{sbact}). Notice that the
$b$ field has become a propagating boson, with the same terms in
the action as a microscopic canonical boson: here these terms
arise from a ($b$,$f$) fermion polarization loop integrated well
away from the $f$ Fermi surface. The parameters $\mu_b, m_b$ may
be interpreted as the chemical potential and mass of the bosons
respectively.
The ($b$,$f$) fermion loop will also lead to higher time and spatial gradient
terms as well as a density-density coupling between $b$ and $f$
in (\ref{sbcrt}), but all these are
formally irrelevant near the quantum critical point of interest.

A key feature of (\ref{sbcrt}), induced by taking the spatial and
temporal continuum limit, is that we have lost information on the
compactness of the U(1) gauge field $a$, {\em i.e.}, the continuum
action is now no longer periodic under $a_{rr'} \rightarrow
a_{rr'} + 2\pi$, as was the case for the lattice action
(\ref{sbact}). The U(1) gauge field is now effectively
non-compact, and consequently monopole excitations have been
suppressed. The monopole gap is finite in
 the U(1) FL$^*$ phase (which is the analog
of the ``Coulomb'' phase of the compact gauge theory) \cite{fs}.
In the FL phase, the monopoles do not exist -- they are confined
to each other.
This occurs due to the condensation of the boson field.
However, the monopole gap
is not expected to close at the transition~\cite{foot4},
and so neglecting the compactness of the gauge field is
permissible.
Indeed, the continuum action (\ref{sbcrt}) provides a
satisfactory description of the critical properties of the FL to
U(1) FL$^*$ transition at $T=0$. However, as we noted in the
caption of Fig~\ref{cross}, the compactness of the gauge field is
crucial in understanding the absence of a $T>0$ phase transition
above the FL phase \cite{compact}.

The action in Eq. (\ref{sbcrt}) above is similar to that popular in
gauge theory descriptions \cite{ki,nagaosa} of the normal state of
optimally doped cuprates but with some crucial differences.
Here the chemical potential of the bosons is fixed while in
Refs.~\onlinecite{ki,nagaosa} the boson density was fixed; as we
will see below, this significantly modifies the physical
implications of the critical theory, and the nature of the
non-Fermi liquid critical singularities as $T>0$. Furthermore, we
are interested specifically in $d=3$, as opposed to the $d=2$ case
considered in Refs.~\onlinecite{ki,nagaosa}.

The phase diagram of the action (\ref{sbcrt}) was sketched in
Fig~\ref{cross}. The horizontal axis, represented in
Fig~\ref{cross} by $J_K$, is now accessed by varying $\mu_b$.
Without any additional (formally irrelevant) second-order time
derivative terms for $b$ in the action, the quantum critical point
between the FL and U(1) FL$^*$ phases occurs precisely at $\mu_b=0
$, $T=0$. We will now discuss the physical properties in the
vicinity of this critical point first at $T=0$, and then at $T>0$,
followed by an analysis of transport properties using the quantum
Boltzmann equation in Section~\ref{sec:qbe}. The final
subsection~\ref{sec:sdworder} will comment on the effect of the
SDW or SDW$^*$ phases that may appear at very low temperatures
(these are not shown in Fig.~\ref{cross}, but sketched in
Fig.~\ref{qcusdw}).

\subsection{Zero temperature}
\label{zerot}

In a mean-field analysis of (\ref{sbcrt}), we see that the FL$^*$
phase (the ``Coulomb'' phase of the gauge theory) obtains for
$\mu_b < 0$ with $\langle b \rangle = 0$, while the FL phase (the
``Higgs'' phase of the gauge theory) obtains for $\mu_b > 0$.

Consider fluctuations for $\mu_b < 0$ in the FL$^*$ phase. Here,
there are no bosons in the ground state, and all self-energy
corrections associated with the quartic coupling $u$ vanish
\cite{fwgf}. The gauge field propagator is given by (\ref{dprop}),
and this does contribute a non-zero boson self energy. At small
momenta $p$ and imaginary frequencies $\epsilon$, the boson
self-energy has the structure (determined from a single
gauge-boson exchange process, as in Refs.~\onlinecite{ki,nagaosa})
\begin{equation}
\Sigma_b (k, i\epsilon) \sim k^2 (1 + c_1 |\epsilon| \ln
(1/|\epsilon|) + \ldots), \label{s1}
\end{equation}
where $c_1$ is some constant. Apart from terms which renormalize
the boson mass $m_b$, these self-energy corrections are less
relevant than the bare terms in the action, and so can be safely
neglected near the critical point. Notice also that $\Sigma_b
(0,0)=0$, and so the quantum critical point remains at $\mu_b =0$.

The critical exponents can now be determined as in
Refs.~\onlinecite{book,fwgf}, and are simply those of the
mean-field theory of (\ref{sbcrt}):
\begin{equation}
\nu=1/2~~~;~~~z=2~~~;~~~\eta=0. \label{critical}
\end{equation}
As in (\ref{s1}) we can also determine the fate the boson
quasiparticle pole as influenced by the gauge fluctuations; we
obtain
\begin{equation}
\mbox{Im} \Sigma_b \left( k, \epsilon=\frac{k^2}{2m_b} \right)
\sim \mbox{sgn} (\epsilon) \epsilon^2 \ln(1/|\epsilon|).
\label{s2}
\end{equation}
The boson lifetime is clearly longer than its energy, and this
pole remains well defined. Finally, we recall our statement in
Section~\ref{sec:specific} that the gauge fluctuations lead to a
$T \ln (1/T)$ specific heat in the FL$^*$ phase, with a diverging
$\gamma$ co-efficient. This behavior remains all the way up to,
and including, the critical point.
Parenthetically, we note that the same calculation in $d=2$
dimensions will yield $C \propto T^{2/3}$.

We turn next to $\mu_b > 0$, in the FL phase. Here the bosons are
condensed, and (\ref{s1}) or explicit calculations show that
\begin{equation}
\langle b \rangle \equiv b_0 \sim \left(\mu_b \right)^{1/2} \sim
\left(J_K - J_{Kc}\right)^{1/2} ,
\end{equation}
where $J_{Kc}$ is the position of the critical point in
Fig~\ref{cross}. The transverse gauge field propagator may be
obtained as in Section~\ref{sec:specific} by integrating out both
the bosons and fermions and expanding the resulting action to
quadratic order; the boson condensate leads to a ``Meissner'' term
in the gauge propagator so that (\ref{dprop}) is replaced by
\begin{eqnarray}
D_{ij}(\vec k, i \omega_n) &\equiv&
\langle a_i(\vec k, i \omega_n)a_j(-\vec k, -\omega_n)\rangle \nonumber \\
&=& \frac{\delta_{ij} - k_ik_j/k^2}{\Gamma |\omega_n|/k + \chi_f
k^2 + \rho_s} \,. \label{dprop2}
\end{eqnarray}
Here $\rho_s$ is the boson ``superfluid density'', and we have
$\rho_s \sim b_0^2$. The presence of such a Meissner term cuts off
the singular gauge fluctuations. The divergence of the specific
heat coefficient $\gamma(T)$ as a function of temperature {\em at}
the critical point implies that it diverges at $T=0$ on
approaching the transition from the FL side. As shown in Appendix
\ref{heatapp}, this is indeed the case, and we find that $\gamma$
diverges as $\gamma \sim \ln(1/b_0)$. In experiments, such a
diverging $\gamma$ is sometimes interpreted as a diverging
effective mass. Importantly, the divergence of $\gamma$ is
unrelated to the singularity in the quasiparticle residue on the
``hot'' Fermi sheet, $Z$, which obeys $Z \sim b_0^2$ as shown in
(\ref{Zhot}), and so vanishes linearly as a function of $J_K -
J_{Kc}$.

\subsection{Non-zero temperatures}
\label{pd}

A crucial change at $T>0$ is that it is now no-longer true that
$\Sigma_b (0,0)=0$ in a region with $\langle b \rangle = 0$.
Instead, as in earlier studies of the dilute Bose
gas~\cite{weichmann,book}, we have
\begin{eqnarray}
\Sigma_b (0,0) &=& 2 u \int \frac{{\rm d}^d k}{(2 \pi)^d} \frac{1}{\exp
\left[k^2
/(2 m_b T)\right] - 1} \nonumber \\
&=& u \frac{\zeta (3/2)}{4 \pi^{3/2}} \left( 2 m_b T\right)^{3/2}
~~~\mbox{in $d=3$}\label{sigmab}
\end{eqnarray}
This behavior determines the crossover phase boundaries shown in
Fig~\ref{cross}. The physical properties are determined by the
larger of the two ``mass'' terms in the $b$ Green's function,
$|\mu_b|$ or $\Sigma_b(0,0)$ -- consequently, the crossover phase
boundaries in Fig.~\ref{cross} lie at $T \sim |\mu_b|^{2/3} \sim
|J_K - J_{Kc}|^{2/3}$. These boundaries separate the U(1) FL$^*$
region at low $T$ and $\mu_b < 0$, and the FL region at low $T$
and $\mu_b > 0$, from the intermediate quantum critical region.
Note that there is no phase transition in the FL region at $T>0$:
this is due to the compactness of the underlying U(1) gauge
theory, and the fact that the ``Higgs'' and ``confining'' phases
are smoothly connected in a compact U(1) gauge theory in three
total dimensions \cite{compact}.

We now briefly comment on the nature of the electrical transport
in the three regions of Fig~\ref{cross}. The behavior is quite
complicated, and we will first highlight the main results by
simple estimates in the present subsection. A more complete
presentation based upon the quantum Boltzmann equation appears in
Section~\ref{sec:qbe}.

The conventional FL region is the simplest, with the usual $T^2$
dependence of the resistivity---the gauge fluctuations are
quenched by the ``Meissner effect''.

In the U(1) FL$^*$ region,
there is an exponentially small density of thermally excited $b$
quanta, and so the boson conductivity $\sigma_b$ is also
exponentially small. As in earlier work \cite{il}, the resistances
of the $b$ and $f$ quanta add in series, and so the total $b$ and
$f$ conductivity remains exponentially small. The physical
conductivity is therefore dominated by that of the $c$ fermions,
which again has a conventional $T^2$ dependence.

Finally, we comment on the transport in the quantum critical
region. This we will estimate following the method of
Ref.~\onlinecite{ki}, with a more complete calculation appearing
in the following subsection. A standard Fermi's Golden rule
computation of scattering off low-energy gauge fluctuations shows
that a boson of energy $\epsilon$ has a transport scattering rate
\begin{equation}
\frac{1}{\tau_{b {\rm tr}} (\epsilon) } \sim T \sqrt{\epsilon}
\end{equation}
for energies $\epsilon \ll T^{2/3}$. From this, we may obtain the
boson conductivity by inserting in the expression
\begin{equation}
\sigma_b \sim \int {\rm d}^3 k \, \tau_{b {\rm tr}} (\epsilon_{bk} ) k^2
\left(- \frac{\partial n (\epsilon_{bk})}{\partial \epsilon_{bk}}\right)
\label{sint}
\end{equation}
where $n (\epsilon) = 1/(e^{\epsilon/T} - 1)$ is the Bose function,
and $\epsilon_{bk} = k^2 /(2m_b) + \Sigma_b (0,0) = k^2/(2 m_b ) +
c_2 T^{3/2}$ for some constant $c_2$. Estimating the integral in
(\ref{sint}) we find that there is an incipient logarithmic
divergence at small $k$ which is cutoff by $\Sigma_b (0,0) \sim
T^{3/2}$, and so $\sigma_b$ diverges logarithmically with $T$:
\begin{equation}
\sigma_b \sim \ln (1/T). \label{slog}
\end{equation}
There are no changes to the estimate of the $f$ conductivity from
earlier work \cite{ki,nagaosa}, and we have $\sigma_f \sim
T^{-5/3}$.
Using again the composition rule of Ref.~\onlinecite{il},
we see that the asymptoptic low-temperature physical conductivity is
dominated by the behavior in (\ref{slog}).

As an aside, we note that for the theory (\ref{sbcrt})
in {\em two} spatial dimensions the result of Eq.~(\ref{slog})
continues to hold, whereas the fermion part becomes
$\sigma_f \sim T^{-4/3}$.
This implies that the asymptoptic low-$T$ physical conductivity
is dominated by (\ref{slog}) in $d=2$ as well.

\subsection{Quantum Boltzmann equation}
\label{sec:qbe}

We now address electrical transport properties of the theory
(\ref{sbcrt}) in more detail, using a quantum Boltzmann equation.
The analysis is in the same spirit as the work of
Ref.~\onlinecite{klw} but, as we have discussed in
Section~\ref{summary}, the variation in the boson density as a
function of temperature leads to very different physical
properties, and requires a distinct analysis of the transport
equation.

We saw in Section~\ref{pd} that the electrical conductivity was
dominated by the $b$ boson contribution, and so we focus on the
time ($t$) dependence described by the distribution function
\begin{equation}
f(\vec{k},t) = \langle b_k^{\dagger} (t) b_k (t) \rangle
\end{equation}
In the absence of an external (physical) electric field $\vec{E}$,
we have the steady state value $f(\vec{k},t) = f_0 (k)$ with
\begin{equation}
f_0 (k) \equiv \frac{1}{\exp \left[(k^2/(2m_b) - \mu_b + \Sigma_b
(0,0))/T\right] - 1}, \label{f0}
\end{equation}
with $\Sigma_b (0,0)$ given in (\ref{sigmab}). The transport
equation in the presence of a non-zero $\vec{E} (t)$ can be
derived by standard means, and most simply by an application of
Fermi's golden rule. The bosons are assumed to scatter off a
fluctuating gauge field with a propagator given by (\ref{dprop})
or (\ref{dprop2}), and this yields the equation
\begin{eqnarray}
&& \frac{\partial f(\vec{k},t)}{\partial t} + \vec{E} (t) \cdot
\frac{\partial f(\vec{k},t)}{\partial \vec{k}} = \nonumber \\
&&- \int_{-\infty}^{\infty} \frac{{\rm d}\Omega}{\pi} \int
\frac{{\rm d}^d q}{(2 \pi)^d}
\mbox{Im} \left[\frac{ k_i D_{ij} (\vec q, \Omega)
k_j }
{m_b^2} \right] \nonumber \\
&&\times (2 \pi) \delta\left( \frac{k^2}{2m_b} -
\frac{(\vec{k}+\vec{q})^2}{2 m_b} - \Omega \right) \nonumber \\
&&~~~~\times \Biggl[ f(\vec{k},t) (1 + f (\vec{k}+\vec{q},t))(1 +
n(\Omega)) \nonumber \\
&&~~~~~~~~~~~~~~~~- f(\vec{k}+\vec{q},t) (1 + f(\vec{k},t))n
(\Omega) \Biggr] \label{qbe}
\end{eqnarray}
where $n(\Omega)$ is the Bose function at a temperature $T$ as above.

\begin{figure}[t]
\epsfxsize=3.1in
\centerline{\epsffile{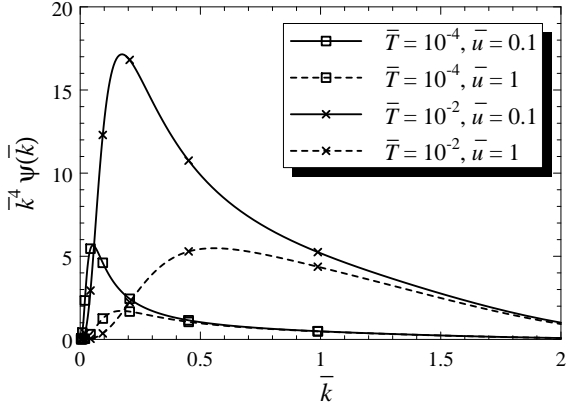}}
\caption{
Plot of the function $\bar{k}^4 \psi(\bar{k})$
for a few values of the reduced temperature $\bar{T}$
and the interaction parameter $\bar{u}$ (\protect\ref{redparm}).
$\psi(\bar{k})$ is defined in Eqs.~(\protect\ref{f0f1}) and
(\protect\ref{psidef}), and has been obtained from the numerical solution
of the quantum Boltzmann equation (\protect\ref{qbefin}).
}
\label{fig:psi}
\end{figure}

We will now present a complete numerical solution of (\ref{qbe})
for the case of a weak, static electric field, to linear order in
$\vec{E}$. The analysis near the quantum critical point parallels
that of Ref.~\onlinecite{ds}, with the main change being that
instead of the critical scattering appearing from the boson
self-interaction $u$, the dominant scattering is from the gauge
field fluctuations (note, however, that it is essential to include
the interaction $u$ to first order in the self-energy shift in
(\ref{f0})). We write
\begin{equation}
f(\vec{k},t) = f_0 (k) + \vec{k} \cdot \vec{E} f_1 (k),
\label{f0f1}
\end{equation}
where notice that $f_1$ depends only on the modulus of $k$ and is
independent of $t$. We now have to insert (\ref{f0f1}) into the
transport equation (\ref{qbe}) and the expression for the
electrical current
\begin{equation}
\vec{J} (t) = \int \frac{{\rm d}^d k}{(2 \pi)^d} \frac{\vec{k}}{m_b}
f(\vec{k},t),
\end{equation}
linearize everything in $\vec{E}$, and so determine the
proportionality between $\vec{J}$ and $\vec{E}$.

\begin{figure}[t]
\epsfxsize=3.1in
\centerline{\epsffile{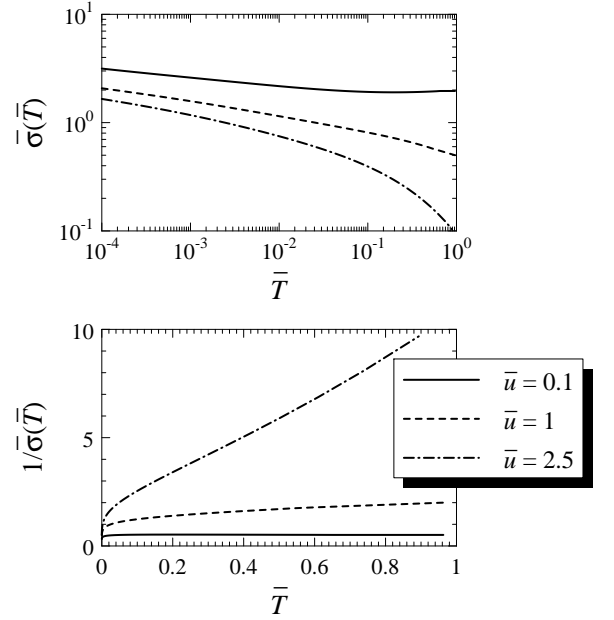}}
\caption{
Scaling function for the boson conductivity, $\bar{\sigma} = \sigma_b / (m_b \chi_f)$,
as function of the reduced temperature $\bar{T}$ for different
values of the interaction parameter $\bar{u}$ (\protect\ref{redparm}).
The results are obtained from the numerical solution
of the quantum Boltzmann equation (\protect\ref{qbefin}) together
with (\protect\ref{sigfin}).
Top panel:  conductivity $\bar\sigma(\bar{T})$ on a log-log scale.
Bottom panel: resistivity $1/\bar\sigma(\bar{T})$ on a linear scale.
}
\label{fig:sig}
\end{figure}

It is useful to re-write the equations in dimensionless
quantities
$\bar{\Omega} = \Omega/T$,
$\bar{k} = k / \sqrt{2 m_b T}$,
$\bar{\sigma} = \sigma_b / (m_b \chi_f)$.
Then it is easy to see that the solution of the quantum Boltzmann
equation at the critical coupling, $\mu_b = 0$,
is characterized by two parameters,
\begin{eqnarray}
\bar{T} = \frac{\chi_f^2}{\Gamma^2} \,(2m_b)^3 \, T \,, \nonumber\\
\bar{u} = u \, \frac{\zeta (3/2)}{4 \pi^{3/2}}\, \frac{\Gamma}{\chi_f} \,,
\label{redparm}
\end{eqnarray}
where $\bar{T}$ is a reduced temperature, and $\bar{u}$
parametrizes the temperature dependence of the effective ``mass'' of
the bosons from Eq. (\ref{sigmab});
$\Gamma$ and $\chi_f$ are the parameters of the gauge
propagator (\ref{dprop}).
The linearized form of the Boltzmann equation (\ref{qbe}) for the function
\begin{equation}
f_1(k) \equiv \psi(k/\sqrt{2 m_b T})
\label{psidef}
\end{equation}
is obtained as
\begin{eqnarray}
- f_0'(\bar{k}) = \int_0^\infty \!\!\!\!\! {\rm d}\bar{k}_1
\left[
K_1(\bar{k},\bar{k}_1) \psi(\bar{k}) + K_2(\bar{k},\bar{k}_1) \psi(\bar{k}_1)
\right]
\label{qbefin}
\end{eqnarray}
with
$f_0'(x) = \partial / (\partial x^2) [\exp(x^2+\bar{u}\sqrt{\bar{T}})-1]^{-1}$;
the expressions for the functions $K_{1,2}$ are given in
Appendix \ref{qbeapp}. From the solution of Eq. (\ref{qbefin}) one obtains
the conductivity according to
\begin{equation}
\bar{\sigma}(\bar{T},\bar{u}) = \frac{1}{6 \pi^2 \sqrt{\bar{T}}}
\int_0^\infty \!\! {\rm d}\bar{k} \, \bar{k}^4\, \psi(\bar{k}) \,.
\label{sigfin}
\end{equation}

The integral equation (\ref{qbefin}) was solved by straightforward
numerical iteration on a logarithmic momentum grid.
We show sample solutions for the function $\bar{k}^4\, \psi(\bar{k})$
in Fig.~\ref{fig:psi}.
The final results for the scaling function of the conductivity are
displayed in Fig.~\ref{fig:sig}.
For small temperatures, the logarithmic divergence of $\sigma_b(T)$
announced in Eq.~(\ref{slog}) is clearly seen;
for larger temperatures the conductivity is exponentially
suppressed due to the temperature-dependent boson mass.
In the crossover region, the results could be fitted with
a power law over a restricted temperature range of roughly one
decade, however, no extended power-law regime emerges.
In comparison with experiments, one has to keep in mind
that the physical resistivity is given by a sum of boson
and fermion resistivities, and that the logarithmically
decreasing low-temperature part of $1/\sigma_b(T)$ cannot be easily
distinguished from a residual resistivity arising from impurities.

\subsection{SDW order}
\label{sec:sdworder}

Our discussion so far has focused primarily on the crossover
between the FL to U(1) FL$^*$ phases, as this captures the primary
physics of the Fermi volume changing transition. At low $T$, we
discussed in Section~\ref{sec:magins} that the longitudinal part
of the gauge fluctuations may induce SDW order on the spinon Fermi
surface of the FL$^*$ phase (leading to the SDW$^*$ phase). On the
FL side of the transition, the gauge fluctuations are formally
gapped by the Anderson-Higgs mechanism. They will, however, still
mediate a repulsive (though finite ranged) interaction between the
quasiparticles at the hot Fermi surface. Furthermore the shape of
the hot Fermi surface evolves smoothly from the spinon Fermi
surface if the FL$^*$ phase. Consequently, it is to be expected
that the SDW order will continue into the FL region up to some
distance away from the transition. Thus it seems unlikely that
there will be a direct transition from SDW$^*$ to FL at zero
temperature. The actual situation then has some similarities to
the mean-field phase diagram in Fig~\ref{pht1}. However,
fluctuations will strongly modify the positions of the phase
boundaries, and we expect that the U(1) FL$^*$ region actually
occupies a larger portion of the phase diagram. Also there is no
sharp transition between the FL and U(1) FL$^*$ regions (unlike
the mean-field situation in Fig~\ref{pht1}), and there is instead
expected to be a large intermediate quantum-critical region as
shown in Fig. \ref{qcusdw}.

%%%%%%%%%%%%%%%%%%%%%%%%%%%%%%%%%%%%%%%%%%%%%%%%%%%%%%%%%%%%%%%%%%

\section{Experimental probes of the U(1) SDW$^*$ state}
\label{sdwstar}

In this Section, we discuss experimental signatures of the U(1)
SDW$^*$ phase focusing particularly on the distinctions with more
conventional SDW metals.

We begin by considering a U(1) SDW$^*$ phase in which a portion of
the spinon Fermi surface remains intact. As discussed in Section
\ref{fluct}, the coupling between the gapless spinons and the
gauge field leads to singularities in the low-temperature
thermodynamics in this phase. In particular the specific heat
behaves as $C(T) \sim T\ln(1/T)$ at low temperature. Thus this
phase is readily distinguished experimentally from a conventional
SDW. Electrical transport in this U(1) SDW$^*$ phase will be
through the conduction electrons with no participation from the
spinons. Thus electrical transport will be Fermi liquid-like. In
contrast thermal transport will receive contributions from both
the conduction electrons and the gapless spinons. Consequently the
thermal conductivity will be in excess of that expected on the
basis of the Wiedemann-Franz law with the free electron Lorenz
number.

The distinction with conventional SDW phases is much more subtle
for U(1) SDW$^*$ phases where the spinons have a full gap. In this
case, there is a propagating gapless linear dispersing photon
which is sharp. The presence of these gapless photon excitations
potentially provides a direct experimental signature of this
phase. It is  extremely important to realize that the emergent
gauge structure of a fractionalized phase is completely robust to
all local perturbations, and is not to be confused with any modes
associated with broken symmetries. Thus despite its gaplessness
the photon is not a Goldstone mode.
% associated with any broken symmetry.
In fact, the gaplessness of the photon is {\em
protected\/}  even if there are small terms in the microscopic
Hamiltonian that break global spin rotation invariance. Being
gapless with a linear dispersion, the photons will contribute a
$T^3$ specific heat at low $T$ which will add to similar
contributions from the magnons and the phonons of the crystal
lattice. In addition, the conduction electrons will contribute a
linear $T$ term. The phonon contribution is presumably easily
subtracted out by a comparison between the heavy Fermi liquid and
magnetic phases. To disentangle the magnon and photon
contributions, it may be useful to exploit the robustness of the
photons to perturbations. Thus for materials with an easy-plane
anisotropy, application of an in-plane magnetic field will gap out
the single magnon, but the photon will stay gapless and will
essentially be unaffected (at weak fields). Thus careful
measurements of field-dependent specific heat may perhaps be
useful in deciding whether the U(1) SDW$^*$ phase is realized.

Finally, quasi-elastic Raman scattering has been suggested as a
probe of the U(1) gauge field fluctuations \cite{nl2} in the
context of the cuprates---the same prediction applies essentially
unchanged here to the fractionalized phases in $d=3$.

Conceptually the cleanest signature of the U(1) SDW$^*$ phase would be detection
of the gapped monopole.
However at present we do not know how this may be directly done in experiments.
Designing such a
``monopole detection'' experiment is an interesting open problem.

%%%%%%%%%%%%%%%%%%%%%%%%%%%%%%%%%%%%%%%%%%%%%%%%%%%%%%%%%%%%%%%%%%

\section{Discussion}
\label{disc}

The primary question which motivates this paper is how to
reconcile a weak moment magnetic metal with non-Fermi liquid
behavior close to the transition to the Fermi liquid. We have
explored one concrete route toward such a reconciliation.
The U(1) SDW$^*$ magnetic states discussed in this paper may be dubbed
spin-charge separated spin density wave metals.
They constitute a class distinct from both the conventional
spin density wave metal and the local-moment metal mentioned in
the Introduction.
However, they
share a number of similarities with both conventional metals. Just
as in the conventional local-moment metal, in the U(1) SDW$^*$
state the local moments do not participate in the Fermi surface.
Despite this the ordering moment may be very small. Indeed this
state may be viewed as a spin density wave that has formed out of
a parent non-magnetic metallic state with a ``small Fermi
surface''. This parent state is a fractionalized Fermi liquid in
which the local moments have settled into a spin liquid and
essentially decoupled from the conduction electrons. The spinons
of the spin liquid form a Fermi surface which undergoes the SDW
transition -- this transition does not affect the deconfinement property
of the gauge field, because the SDW order parameter is gauge neutral and
thus does not effectively couple to the gauge field excitations.

We showed that in the region of evolution from this state to the
conventional Fermi liquid, non-Fermi liquid behavior obtains (at
least at intermediate temperatures). We also argued that the
underlying transition that leads to this non-Fermi liquid physics
is the Fermi volume changing transition from FL to FL$^*$. Despite
the jump in the Fermi volume, this transition is continuous and
characterized by the vanishing of the quasiparticle residue $Z$ on
an entire sheet of the Fermi surface (the ``hot'' Fermi surface)
on approaching the transition from the FL side.

A specific heat that behaves as $T\ln(1/T)$ is commonly observed
in a variety of heavy-fermion materials close to the transition to
magnetism. In the context of the ideas explored in this paper,
such behavior of the specific heat is naturally obtained in
{\em three-dimensional} systems.
A small number of heavy-fermion
materials exhibit such a singular specific heat even in the
presence of long-ranged magnetic order. As we have emphasized,
precisely such non-Fermi liquid specific heat obtains in one of
the exotic magnetic metals discussed in this paper (the U(1)
SDW$^*$ phase with a partially gapped spinon Fermi surface). It
would be interesting to check for violations of the
Wiedemann-Franz law at low temperature in such materials.

A general point emphasized in this paper is that the observed
non-Fermi liquid physics near the onset of magnetism actually has
little to do with fluctuations of the magnetic order parameter.
Rather we propose that the non-Fermi liquid physics is associated
with the destruction of the large Fermi surface. The concrete
realization of this picture explored in this paper is
that the destruction of the large Fermi surface leads to a
fractionalized Fermi liquid which eventually (at low temperature)
develops spin density wave order. As we discussed extensively, the
resulting spin density wave state is an exotic magnetic metal.

It is also of interest to consider a different scenario in which the
small Fermi surface state is unstable at low temperature toward
{\em confinement} of spinons and magnetic order.
It is particularly interesting to consider such a scenario in $d = 2$.
The physics of the Fermi-volume changing fluctuations is again described
by a theory of condensation of a slave boson field coupled to a Fermi surface of
spinons by a U(1) gauge field.
For a non-compact U(1) gauge field, such a theory has a number of interesting
properties. As noticed by Altshuler {\em et al.}~\cite{altshuler}, the spin
susceptibility at extremal
wavevectors of the spinon fermi surface have (possibly divergent) singularities
due to the gauge fluctuations.
Indeed the spin physics
of this model is critical and described by a non-trivial fixed point. The
dynamical susceptibility
at these extremal wavevectors and at a frequency $\omega$ is expected to satisfy
$\omega/T$ scaling.
For a general spinon Fermi surface these extremal wavevectors will chart out
one-dimensional lines in the Brillouin zone
at which critical scattering will be seen in inelastic neutron scattering.
A spin density wave instability can develop out of this critical state at a
particular extremal wavevector where
the amplitude of the diverging susceptibility is the largest.
Arguments very similar to those in Section \ref{pd} also
show that transport will be governed by non-Fermi liquid power laws in this
theory.

There is a striking {\em qualitative} resemblance between these results and
the experiments on CeCu$_{6-x}$Au$_x$.
At the critical Au concentration neutron scattering experiments see critical
scattering
on lines in the Brillouin zone satisfying $\omega/T$ scaling \cite{schroeder}.
Magnetic ordering occurs at particular wavevectors on this line.
Furthermore, empirically the spin fluctuations appear to be quasi
two-dimensional, suggesting that the ideas sketched above may indeed be
relevant.
We note that they significantly differ from earlier proposals to explain the
behavior of CeCu$_{6-x}$Au$_x$ \cite{cecuau}.
As mentioned in the text, the specific heat in the $d=2$ quantum critical region
will have the form $C/T \sim T^{-1/3}$; interestingly, such a behavior has been
observed in YbRh$_2$Si$_2$ in the low-temperature regime near a
quantum-critical point~\cite{ybrhsi}.
On the theoretical front, there are a number of conceptual issues~\cite{sp}
related to the legitimacy of ignoring the compactness of the gauge field in $d =
2$.
Developing a more concrete theoretical description of these general ideas is an
interesting challenge for future work.

\begin{acknowledgments}
We thank P.~Coleman, M.P.A. Fisher, E.~Fradkin, A.~Georges, L.~Ioffe, Y.-B.~Kim,
G.~Kotliar, A.~Millis, N.~Prokof'ev, T.~M.~Rice, Q.~Si, M.~Sigrist,
A.~Tsvelik, and X.-G. Wen for useful discussions. T.S is particularly grateful
to
Patrick Lee for a number of enlightening conversations that
clarified his thinking. This research was supported by the MRSEC
program of the US NSF under grant number DMR-0213282 (T.S.), by US
NSF Grant DMR 0098226 (S.S.), and by the DFG Center for Functional
Nanostructures at the University of Karlsruhe (M.V.). T.S. also
acknowledges funding from the NEC Corporation, and the Alfred P.
Sloan Foundation and the hospitality of Harvard University where part of this
work was done.
\end{acknowledgments}

%%%%%%%%%%%%%%%%%%%%%%%%%%%%%%%%%%%%%%%%%%%%%%%%%%%%%%%%%%%%%%%%%%

\appendix

\section{Oshikawa's argument and topological order}
\label{oshi}

Oshikawa has presented~\cite{oshi2} an elegant non-perturbative
argument demonstrating that the volume of the Fermi surface is
determined by the total number of electrons in the system. In our
previous work \cite{ssv1}, and in the present paper, we have
argued for the existence of a non-magnetic FL$^*$ state with a
different Fermi surface volume. As we discussed earlier
\cite{ssv1}, this apparent conflict is resolved when we allow for
global topological excitations in Oshikawa's analysis; such
excitations emerge naturally in the gauge theories we have
discussed for the FL$^*$ state. In other words, Oshikawa's
argument implies that violation of Luttinger's theorem must be
accompanied by topological order.

In this Appendix, we briefly recall the steps in Oshikawa's
argument, and show how it can be modified to allow for a small
Fermi surface in a FL$^*$ state. As far as possible, we follow the
notation of Oshikawa's paper~\cite{oshi2}.

For definiteness, consider a two-dimensional Kondo lattice with a
unit cell of lengths $a_{x,y}$. The ground state is assumed to be
non-magnetic, with equal numbers of up and down spin electrons.
Place it on a torus of lengths $L_{x,y}$, with $L_x/a_x$,
$L_y/a_y$ co-prime integers. Adiabatically insert a flux $\Phi = 2
\pi$ ($\hbar=c=e=1$) into one of the holes of the torus (say the
one enclosing the $x$ circumference), acting only on the up-spin
electrons. Then the initial and final Hamiltonians are related by
a unitary transformation generated by
\begin{equation}
U = \exp \left( \frac{2 \pi i}{L_x} \sum_{r} n_{r \uparrow}
\right)
\end{equation}
where $n_{r \uparrow}$ is the number operator of all electrons
(including the local moments) with spin up on the site $r$. After
performing the unitary transformation to make the final
Hamiltonian equivalent to the initial Hamiltonian, the final and
initial states are found to have a total crystal momentum which
differs by
\begin{equation}
\Delta P_x = \frac{2 \pi}{L_x} \frac{L_x L_y}{v_0}
\frac{\rho_a}{2} \left( \mbox{mod} \frac{2\pi}{a_x} \right)
\label{o1}
\end{equation}
where $v_0 = a_x a_y$ is a volume of a unit cell, the second
factor on the r.h.s. counts the number of unit cells in the
system, and $\rho_a = 2 \rho_{a\uparrow}$ is the mean number of
electrons in every unit cell. Clearly the crystal momentum is
defined modulo $2 \pi/a_x$, and hence the modulus in (\ref{o1}).

Now imagine computing the change in crystal momentum by studying
the response of the quasiparticles to the inserted flux. As shown
by Oshikawa, the quasiparticles associated with a Fermi surface of
volume $\mathcal{V}$ lead to a change in momentum which is
\begin{equation}
\Delta P^q_x = \frac{2 \pi}{L_x} \frac{\mathcal{V}}{(2 \pi)^2/(L_x
L_y)} \left( \mbox{mod} \frac{2\pi}{a_x} \right), \label{o2}
\end{equation}
where the second factor on the r.h.s. counts the number of
quasiparticles within the Fermi surface. Equating $\Delta P_x$ and
$\Delta P^q_x$, and the corresponding expressions for $\Delta P_y$
and $\Delta P^q_y$, Oshikawa obtained the conventional Luttinger
theorem, which applies to the volume
$\mathcal{V}=\mathcal{V}_{FL}$ of the Fermi surface in the FL
state
\begin{equation}
2 \frac{v_0}{(2 \pi)^2} \mathcal{V}_{FL} = \rho_a \mbox{(mod 2)}
\end{equation}

In the FL$^*$ state, there are additional low-energy excitations
of the local moments that yield an additional topological
contribution to the change in crystal momentum. Indeed, the
influence of an insertion of flux $\Phi$ is closely analogous to
the transformation in the Lieb-Schultz-Mattis~\cite{lsm} argument,
and it was shown~\cite{bonesteel,misguich} that a spin liquid state
in $d=2$ acquires the momentum change
\begin{equation}
\Delta P^t_x = \frac{\pi}{a_x} \frac{L_y}{a_y} \left( \mbox{mod}
\frac{2\pi}{a_x} \right) \label{dpt}
\end{equation}
where the second factor on the r.h.s. now counts the number of
rows which have undergone the Lieb-Schultz-Mattis transformation.
Now using $\Delta P_x = \Delta P^q_x + \Delta P^t_x$, we now
obtain the modified Luttinger theorem obeyed in the FL$^*$ phase:
\begin{equation}
2 \frac{v_0}{(2 \pi)^2} \mathcal{V}_{FL^*} = (\rho_a-1) \mbox{(mod
2)}.
\end{equation}

It is clear that the above argument is easily extended to a $Z_2$
FL$^*$ state in $d=3$. The case of U(1) FL$^*$ state in $d=3$ is
somewhat more delicate because there is now a gapless spectrum of
gauge fluctuations which can contribute to the evolution of the
wavefunction under the flux insertion; nevertheless, the momentum
change in (\ref{dpt}) corresponds to an allowed gauge flux, and we
expect that (\ref{dpt}) continues to apply.

%%%%%%%%%%%%%%%%%%%%%%%%%%%%%%%%%%%%%%%%%%%%%%%%%%%%%%%%%%%%%%%%%%

\section{Toy model with U(1) fractionalization and a spinon Fermi
surface}
\label{toy}

In this Appendix, we will display a concrete model in three
spatial dimensions that is in a U(1) fractionalized phase in
three dimensions, and has a Fermi surface of spinons coupled to a
gapless U(1) gauge field. As discussed earlier, this spinon
Fermi surface could eventually (at low energies) undergo various
instabilities including in particular to a spin density wave
state.

Consider the following model:
\begin{eqnarray}
H & = & H_{t\psi} + H_{\Delta} +H_b +H_u + H_U \,, \\
H_{t\psi} & = & -\sum_{\langle rr'\rangle}t
\left(\psi^{\dagger}_r\psi_{r'} + \mbox{h.c.}\right) \,,\nonumber\\
H_{\Delta} & = & \Delta \sum_{\langle rr'\rangle}e^{i\phi_{rr'}}\left(
\psi^{\dagger}_{r\uparrow}\psi^{\dagger}_{r'\downarrow}-
\psi^{\dagger}_{r'\uparrow}
\psi^{\dagger}_{r\downarrow}\right)+\mbox{h.c.} \,,\nonumber\\
H_b & = & -w\sum_{[rr'r'']} \cos(\phi_{rr'}-\phi_{rr''}) \,,\nonumber\\
H_u & = &u\sum_{\langle rr' \rangle}n_{rr'}^2 \,,\nonumber\\
H_U & = & U\sum_r (N_r-1)^2 \,. \nonumber
\end{eqnarray}
Here $\psi_{r}$ destroys a spinful charge-$1$ electron at each
site of a cubic lattice in three spatial dimensions;
$e^{i\phi_{rr'}}$ creates a charge-$2$, spin-$0$ ``Cooper pair''
that resides on the links. $n_{rr'}$ is conjugate to $\phi_{rr'}$
and may be regarded as the Cooper pair number associated with each
link. $N_r$ is the total charge associated with each site and is
given by
\begin{equation}
N_r  =  \sum_{r' \in r}n_{rr'} + \psi^{\dagger}_r\psi_r.
\end{equation}
The Hamiltonian $H$ may be regarded as describing a system of
electrons coupled with strong phase fluctuations. The first term
in $H_b$ represents Josephson coupling between two ``nearest
neighbor'' bonds. $H_u$ penalizes fluctuations in the Cooper pair
number at each bond. $H_U$ penalizes fluctuations in the total
charge $N_r$ that can be associated with each lattice site.
 The total charge of the full system clearly is
\begin{equation}
N_{\rm tot} = \sum_r N_r.
\end{equation}

Depending on the various model parameters, several distinct phases
are possible. Here we focus on the limit of large $U$.
Diagonalizing $H_U$ requires that the ground state(s) satisfy $N_r
= 1$ at all sites $r$. There is a gap of order $U$ to states that
do not satisfy this condition. Clearly the system is insulating in
this limit.

The condition $N_r = 1$ for all $r$ still allows for a huge
degeneracy of ground states which will be split once the other
terms in the Hamiltonian are included. This splitting may be
described by deriving an effective Hamiltonian that lives in the
space of degenerate states specified by $N_r = 1$. As discussed in
Ref.~\onlinecite{bosfrc}, this effective Hamiltonian may be usefully
viewed as a (compact) U(1) gauge theory. This may be explicitly
brought out in the present case by the change of variables
\begin{eqnarray}
\phi_{rr'} & = & \epsilon_r a_{rr'} \,,\nonumber\\
n_{rr'} & = & \epsilon_r E_{rr'} \\
\psi_{r\alpha} & = & f_{r\alpha}~~ \mbox{for}~~ r \in A \,,\nonumber\\
\psi_{r\alpha} & = & i\sigma^y_{\alpha \beta}f^{\dagger}_{r\beta}
~~\mbox{for}~~ r \in B \,.
\end{eqnarray}
Here $\epsilon_r = +1$ on the $A$ sublattice and $-1$ on the $B$
sublattice. In terms of these variables, the constraint $N_r = 1$
reads
\begin{equation}
    \vec \nabla \cdot \vec E + f^{\dagger}_r f_r = 1
\end{equation}
at each site $r$. (We note that both $\vec a$ and $\vec E$ may be regarded
as vector fields defined on the lattice). At order $w^2/U, u, t,
\Delta$, the effective Hamiltonian takes the form
\begin{eqnarray}
    H_{\rm eff} & = & H_K + H_u + H_f \,,\\
    H_K & = & -K \sum_P \cos(\vec \nabla \times \vec a) \,,\nonumber\\
    H_u & = & + u\sum_r \vec E^2 \,,\nonumber\\
    H_f & = & -\Delta \sum_{\langle rr' \rangle} \left(e^{ia_{rr'}}
    f^{\dagger}_rf_{r'} + \mbox{h.c.} \right) \,.\nonumber
\end{eqnarray}
Here $K =2w^2/U$, and the sum $\sum_P$ runs over elementary
plaquettes.
$H_{\rm eff}$ together with the constraint may be
viewed as a Hamiltonian for a compact U(1) gauge theory coupled
to a gauge charge-$1$ fermionic matter field $f$. (Note that to
leading order the $t$ term does not contribute). $H_{\rm eff}$ still
admits several different phases depending on its parameters. Of
interest to us is the limit $K \sim w^2/U \gg u$. In this limit,
monopoles of the compact U(1) gauge field will be gapped.
Consequently at low energies, we may take the gauge field to be
non-compact. The $\cos(\vec \nabla \times \vec a)$ term can then be
expanded to quadratic order to get the usual Maxwell dynamics for
the gauge field. The $f$-particles form a Fermi surface which is
coupled to this gapless U(1) gauge field.
Note that in the low-energy manifold with $N_r = 1$ at all $r$,
all excitations have
zero physical electric charge. Thus the $f$ particles are neutral
fermionic spinons.

As with any Fermi surface, this spinon Fermi surface state could
at low energies further undergo various instabilities to other
states (density waves, pairing, etc) depending on the residual
interactions between the spinons. There are various sources of
such interactions: First there is the gauge interaction that is
explicit in $H_{\rm eff}$ in the leading order. Then the $t$ term
contributes to $H_{\rm eff}$ at second order and leads to a quartic
spinon-spinon interaction as well. The specific low-energy
instability of the spinon Fermi surface will be determined by the
details of the competition between these various sources of
interaction, and will not be discussed further here for this
model.

Apart from the deconfined phase discussed above, the model
possesses confined phases; for large $U$ those occur for smaller $u$,
and the deconfinement transition occurs through the condensation
of monopoles in the gauge field.

%%%%%%%%%%%%%%%%%%%%%%%%%%%%%%%%%%%%%%%%%%%%%%%%%%%%%%%%%%%%%%%%%%

\section{Mean-field phase diagram in an external Zeeman magnetic field}
\label{mfbext}

\begin{figure}[t]
\epsfxsize=3.2in \centerline{\epsffile{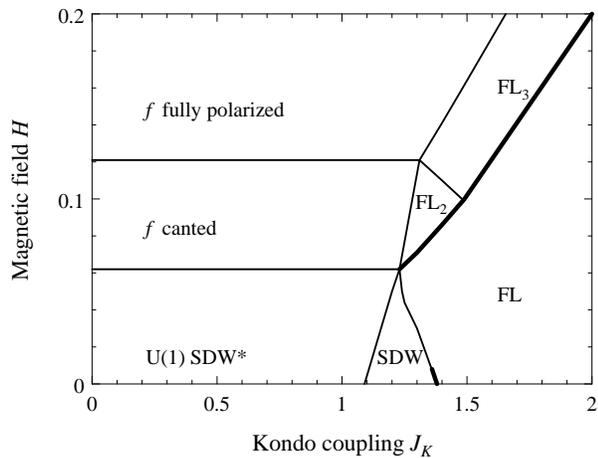}} \caption{
Mean-field phase diagram of $H_{\rm mf}$ (\protect{\ref{mf2}}) on
the cubic lattice, now as function of Kondo coupling $J_K$ and
external field $H_z$ at $T=0$. Parameter values are as in
Fig.~\protect\ref{pht1}. For a description of the phases see text.
} \label{phh1}
\end{figure}

In this Appendix we briefly discuss the behavior of the
mean-field theory of Sec.~\ref{mfmag} in an externally applied
field.
A sample zero-temperature phase diagram is displayed in Fig.~\ref{phh1},
which shows very rich behavior.

The phases at small fields are straightforward generalizations of
the low-temperature zero-field phases of Fig.~\ref{pht1}: The U(1)
SDW$^*$ has weakly polarized conduction electrons, $b_0=0$,
non-zero $\chi_0$ indicating spinon hopping, and a canted spinon
magnetization $\vec M_r$ with a staggered component along $\hat x$
and a uniform component along $\hat z$. The SDW phase has similar
characteristics, but now $b_0\neq 0$ indicating a conventional
weakly field-polarized magnet with confinement. Finally, the FL
phase has $b_0\neq 0$, $\chi_0\neq 0$, weakly polarized heavy
quasiparticles, and the mean-field parameter $\vec{M}_r$ has only
a uniform $z$ component.

In the small-$J_K$ region, increasing external field progressively
suppresses the effect of $J_H$. At intermediate field, a phase
with ``canted'' $f$ moments arises, where now $\chi_0 = b_0 = 0$
(no spinon hopping), and $\vec M_r$ is canted as described above.
Larger fields fully polarize the local moments, i.e., $\vec{M}_r$
points uniformly along $\hat z$ with maximum amplitude, and
$\chi_0 = b_0 = 0$. This phase is also realized for larger $J_K$
and large fields -- here the field quenches the Kondo effect. On
the Fermi-liquid side of the phase diagram, two more phases arise
in the present mean-field theory which are labelled by FL$_2$ and
FL$_3$ in Fig.~\ref{phh1}; both have non-zero $b_0$ and $\chi_0$.
In FL$_2$, the mean-field parameter $\vec{M}_r$ has both staggered
and uniform components, {\em i.e.}, this phase describes canted,
weakly screened local moments. Turning to the FL$_3$ phase, this
high-field phase has the same symmetry characteristics as FL at
intermediate fields, but a different Fermi surface topology.
Whereas FL phase at intermediate fields has a single Fermi surface
sheet for one spin direction (the majority spin have one full and
one empty band whereas the minority spins have one partially
filled and one empty band), in FL$_3$ the upper band of the
majority spins becomes partially filled, too. FL and FL$_3$ are
separated by a strongly first-order transition in mean-field
theory. There are numerous other phase transitions associated with
a change in the Fermi surface topology -- those do not display
strong thermodynamic signatures and are not shown. We note that
for the field range displayed in Fig.~\ref{phh1}, $|{\vec H}_{\rm
ext}| \ll t$, the conduction electrons are in general weakly
affected by the field; significant polarization of them occurs
only at much higher fields.

Notably, smaller values of the decoupling parameter $x$ admit yet
another field-induced transition in the small-$J_K$ region: If the
magnetism is very weak, {\em i.e.}, the spinons have a small gap
compared to their bandwidth, then a small applied field can close
the spinon gap without significantly affecting their band
structure. Such a transition would yield a kink in the
magnetization of the local-moment subsystem as function of the
applied field, implying a ``metamagnetic'' behavior which is here
generically associated with a {\em continuous} transition.

%%%%%%%%%%%%%%%%%%%%%%%%%%%%%%%%%%%%%%%%%%%%%%%%%%%%%%%%%%%%%%%%%%

\section{Specific heat singularity}
\label{heatapp}

Here we present some details on the calculation of the singular
specific heat coming from gauge fluctuations. The calculations in
the FL$^*$ phase and at the critical point are standard. We will
therefore only consider the FL phase. In this phase close to the
critical point, transverse gauge fluctuations are described by the
action
\begin{equation}\label{strgg}
S =
\int\frac{{\rm d}^3 k}{(2\pi)^3}\frac{1}{\beta}\sum_{\omega_n}\left(\frac{|\omega_n|}{k
}
+ k^2 + \rho_s\right)|\vec a(\vec k, \omega_n)|^2.
\end{equation}
As explained in Section \ref{flcrt}, close to the transition
$\rho_s \sim b_0^2$.
This gives a free energy
\begin{equation}\label{ggfe}
  F =
\frac{2}{\beta}\sum_{\omega_n}\int\frac{{\rm d}^3 k}{(2\pi)^3}\ln\left(\frac{|\omega_n|
}{k}
+ k^2 + \rho_s\right).
\end{equation}
To calculate the low-temperature specific heat, we need the change
in free energy on going from zero to a small non-zero temperature.
After a Poisson resummation this is given by
\begin{eqnarray}
  \delta F(T) & \equiv & F(T) - F(0) \\
  & = & 2\int_{\vec k}\sum_{m
  \neq 0}\int\frac{{\rm d}\omega}{2\pi}e^{i\beta m\omega}\ln\left(\frac{|\omega|}{k}
+ k^2 + \rho_s\right) \nonumber\\
& = & 2\int_{\vec k, \omega}\int_0^{\infty} {\rm d}\lambda\sum_{m
  \neq 0}
  \frac{ke^{i\beta m\omega}}{|\omega|
+ k(k^2 + \rho_s + \lambda)}\nonumber \\
& = & 2\int_{\vec k, \omega, \lambda}\sum_{m \neq
0}k\int_{0}^{\infty} {\rm d}u e^{i\beta m\omega-u(|\omega| + k(k^2 + \rho_s +
\lambda))} \nonumber \,.
\end{eqnarray}
The $\omega, \lambda$ integrals may now be performed to obtain
\begin{eqnarray}
\delta F(T) & = & \frac{4}{\pi}\int_{\vec k}\sum_{m = 1}^{\infty}
\int {\rm d}u \frac{e^{-uk(k^2 + \rho_s)}}{u^2 + (m\beta)^2} \nonumber\\
& = &  \int_0^{\Lambda} \frac{{\rm d}k k^2}{2\pi^3}\int_u \frac{(\pi uT
\coth (\pi u T) - 1)e^{-uk(k^2 + \rho_s)}}{u^2} \nonumber .
\end{eqnarray}
In the last equation we have introduced an upper cut-off $\Lambda$
for the momentum integral. The remaining integrals can now be
straightforwardly evaluated for small $T$, and we find
\begin{equation}\label{dff}
  \delta F(T) =
  \frac{T^2}{12\pi}\ln\left(\frac{\Lambda^2}{\rho_s}\right).
\end{equation}
Thus the specific heat
\begin{equation}\label{spht}
  C(T) = \gamma T
\end{equation}
with $\gamma \sim \ln(1/\rho_s) \sim \ln(1/b_0)$. Setting
$\rho_s=0$, a similar calculation also shows that $C(T) \sim T
\ln(1/T)$ in the FL$^*$ phase.

For completeness, we mention the corresponding behavior
in two dimensions. In analogy to the above calculations,
we find $C(T) \propto T^{2/3}$ in the quantum-critical and
FL$^*$ regions.
%, whereas $C/T=\gamma \propto \rho_s^{-1/3}$ at small $T$ in the FL phase.

%%%%%%%%%%%%%%%%%%%%%%%%%%%%%%%%%%%%%%%%%%%%%%%%%%%%%%%%%%%%%%%%%%

\section{Details of the quantum Boltzmann equation}
\label{qbeapp}

In the following we describe a few details of the derivation
of the linearized version of the quantum Boltzmann equation
(\ref{qbefin}) in Sec.~\ref{sec:qbe}.
Inserting the ansatz (\ref{f0f1}) into Eq. (\ref{qbe})
leads to a scalar equation for $f_1$.
The frequency integral is easily performed;
the remaining momentum integral can be split into radial and angular
part.
This directly yields Eq.~(\ref{qbefin}), with
\begin{eqnarray}
K_1(\bar{k},\bar{k}_1) &=&
\left[1+f_0(\bar{k}_1)+n(\bar{k}^2-\bar{k}_1^2)\right] \\ && \times
\frac{\bar{k}\, \bar{k}_1^2}{4\pi^2} \, \int_{-1}^{1} {\rm d}x \,
K\left(x,\frac{\bar{k}_1}{\bar{k}},\bar{k}\sqrt{\bar{T}}\right) \, , \nonumber
\end{eqnarray}

\begin{eqnarray}
K_2(\bar{k},\bar{k}_1) &=&
\left[f_0(\bar{k})-n(\bar{k}^2-\bar{k}_1^2)\right] \\ && \times
\frac{\bar{k}\, \bar{k}_1^2}{4\pi^2} \, \int_{-1}^{1} {\rm d}x \,\frac{x \bar{k}_1}{\bar{k}}
K\left(x,\frac{\bar{k}_1}{\bar{k}},\bar{k}\sqrt{\bar{T}}\right) \nonumber
\end{eqnarray}
with $f_0(x) = [\exp(x^2+\bar{u}\sqrt{\bar{T}})-1]^{-1}$ and
$n(x) = (e^x-1)^{-1}$.
The kernel $K(x,\bar{k}_1/\bar{k},\bar{k}\sqrt{\bar{T}})$ is given by
\begin{eqnarray}
K(x,\alpha,\lambda) &=& {\rm Im}\,\left(
-i\frac{1-\alpha^2}{\sqrt{\alpha^2\!+\!1\!-\!2\alpha x}} + \lambda(\alpha^2\!+\!1\!-\!2\alpha x)
\right)^{-1} \nonumber\\
&& ~~\times \left(
1-\frac{(\alpha x-1)^2}{\alpha^2+1-2\alpha x}
\right) \,.
\nonumber
\end{eqnarray}
The integrals necessary for the evaluation of $K_{1,2}$ are of the form
\begin{eqnarray}
\int {\rm d}y \, \frac{y^{n-1/2}}{y^3+C}
\nonumber
\end{eqnarray}
with $n=0,\ldots,3$ and can be performed analytically.

The numerical solution of Eq.~(\ref{qbefin}) is done by re-writing it in
the form
\begin{eqnarray}
\psi(\bar{k}) &=& - f_0'(\bar{k}) \nonumber \\
&&\times \left( \int_0^\infty \!\!\!\!\! {\rm d}\bar{k}_1
\left[
K_1(\bar{k},\bar{k}_1)  + K_2(\bar{k},\bar{k}_1) \frac{\psi(\bar{k}_1)}{\psi(\bar{k})}
\right]
 \right)^{-1}
\nonumber
\end{eqnarray}
which allows for a stable numerical iteration.

%%%%%%%%%%%%%%%%%%%%%%%%%%%%%%%%%%%%%%%%%%%%%%%%%%%%%%%%%%%%%%%%%%

%\end{multicols}

\begin{thebibliography}{}
\bibitem{foot1} In our paper, we are using the term ``Luttinger's
theorem'' to simply be equivalent to the statement that the volume
enclosed by the Fermi surface counts all electrons. With this
definition, the state proposed by Essler and Tsvelik
\protect\cite{tsvelik} does amount to a violation of ``Luttinger's
theorem''. However, in their case, the violation of the formal
many body proof of Oshikawa can be understood in the context of
renormalized perturbation theory. In our models, the violation is
intimately linked to topological order, and cannot be understood
by any perturbative analysis of the electron Hamiltonian.
\bibitem{foot2}
Quite generally a U(1) deconfined state necessarily requires a
conserved U(1) gauge charge for the spinons. In contrast, in a
state described by a deconfined $Z_2$ gauge theory, consistency
only requires a conserved $Z_2$ charge for the spinons: this
implies that the spinons will generically be paired in $Z_2$
states but not in the U(1) states.
\bibitem{foot3} In particular, the monopole gap will survive a weak Kondo
coupling.
\bibitem{foot4} The fate of the monopoles at this transition is analogous to the
fate of
gapped spinons on going from a spin liquid to a spin-Peierls state:
though the spin gap does not close, the spinons
exist only on one side of the transition and play no role
in the universal low-energy critical properties.
\end{thebibliography}
\end{document}